\documentclass[english]{article}
\usepackage[latin9]{inputenc}
\usepackage{babel}
\usepackage{float}
\usepackage{amsmath}
\usepackage{amssymb}
\usepackage{graphicx}
\usepackage{subcaption}
\usepackage[unicode=true]
 {hyperref}
\usepackage[export]{adjustbox}

\makeatletter

\providecommand{\tabularnewline}{\\}
\floatstyle{ruled}
\newfloat{algorithm}{tbp}{loa}
\providecommand{\algorithmname}{Algorithm}
\floatname{algorithm}{\protect\algorithmname}

\usepackage{hyperref}
\usepackage{algpseudocode}
\usepackage{algorithm}
\usepackage{algorithmicx}
\usepackage{mathrsfs}

\algrenewcommand\algorithmicrequire{\textbf{Input:}}
\algrenewcommand\algorithmicensure{\textbf{Output:}}
\algdef{SE}[DOWHILE]{Do}{doWhile}{\algorithmicdo}[1]{\algorithmicwhile\ #1}%
\usepackage{xcolor}
\makeatother

\begin{document}

\title{Model Order Reduction for Temperature-Dependent Nonlinear Mechanical
	Systems: A Multiple Scales Approach}

\author{Shobhit Jain, Paolo Tiso}
	\date{\vspace{-5ex}}
	\author{Shobhit Jain\footnote{Corresponding author, E-mail: shjain@ethz.ch, Phone: +41 44 632 77 55}, Paolo Tiso}
\maketitle
	\begin{center}
		Institute for Mechanical Systems, ETH Zürich \\
		Leonhardstrasse 21, 8092 Zürich, Switzerland
		\par\end{center}
\begin{abstract}
	The thermal dynamics in thermo-mechanical systems exhibits a much
	slower time scale compared to the structural dynamics. In this work, we use the method of multiple scales to reduce the thermo-mechanical structural models with a slowly-varying temperature distribution in a systematic manner. In the process, we construct a reduction basis that adapts according to the instantaneous temperature distribution of the structure, facilitating an efficient reduction in the number of unknowns. As a proof of concept, we demonstrate the method on a range of linear and nonlinear beam examples and obtain a consistently better accuracy and reduction in the number of unknowns than the standard Galerkin projection using a constant basis.
\end{abstract}

\section{Introduction}
\label{sec:introduction}
Model reduction plays an important role in identifying low dimensional
features in high-dimensional dynamical systems, whose numerical simulation
would otherwise be either extremely computationally-intensive or entirely
infeasible. In such cases, model reduction leads to a small number
of variables that describe the system dynamics evolving over low-dimensional
subspaces (or manifolds), making computations easier. Structural dynamics
applications such as aerospace structures, civil structures, micro-electro-mechanical systems (MEMS)
etc., often involve such high-dimensional dynamical systems, which
are obtained after spatial discretization of the governing equations
of motion. Such structures also exhibit changes in mechanical behavior
under the influence of external environmental effects such as thermal,
acoustic etc., which manifest as coupling terms in the equations of
motion. Model reduction in such coupled domains is especially challenging. We propose a multiple-scales-based technique to reduce structural dynamics equations varying geometrical nonlinear structural dynamics. In this work, we focus on structural systems featuring geometric nonlinearities and thermal stresses induced by temperature fields varying slowly in time and space, and we propose a multiple-scales-based technique to reduce the temperature-dependent equations of motion. 

For modeling mechanical structures under the influence of temperature
changes, a one-way coupling is often used. This means that the temperature
changes affect the structural dynamics but not vice-versa. This is a reasonable assumption because roughly speaking, the thermal dynamics evolves over a much slower time-scale in comparison with the structural dynamics and though the structure can respond to temperature-changes, the same is not true for thermal response to structural motion. More specifically, this slow thermal time scale is reflected in the large time constant of temperature evolution law in comparison with the time-period of oscillation of the structure. Due to such a slow nature of temperature evolution, structural dynamics equations are sometimes even modeled with a constant temperature over a given structural time span. When the temperature does not change during a structural dynamic simulation, reduced-order models (ROM) can be constructed using many standard techniques in literature, e.g., using vibration modes (VMs) \cite{Rixen97}, modal derivatives \cite{IC}, proper orthogonal decomposition (POD) \cite{POD1,POD2}, implicit condensation and expansion (ICE) \cite{Hollkamp_ICE}, among others, cf.~ref.~\cite{mignolet} for a review. 

Nonlinear ROMs are often local in nature, i.e, these are constructed to approximate dynamical behavior locally in the vicinity of an equilibrium, as opposed to globally in the phase space. This is, especially, the case when ROMs are constructed from vibration modes and modal derivatives (cf.~refs.~\cite{IC,qm}), which are computed around an equilibrium and span the relevant subspaces. In the context of temperature dependence, consider a simple linear
mechanical system such as 
\begin{equation}
M\ddot{x}+C\dot{x}+K(T)x + =f(t)+g(T),
\end{equation}
where $M,C,K$\footnote{we introduce these equations/symbols informally for the purpose of the discussion in this introductory section. A more precise treatment follows from Section~\ref{sec:setup} onwards.} are mass, damping and stiffness matrices; $ t $ is the time; and $T$
represents temperature variable(s). Usually, model reduction is performed by assuming the temperature $T$ to be a static parameter and a reduction
basis (comprising, e.g., a few VMs) is computed for
a given value of this static temperature parameter. Interestingly, in specific cases, a basis comprising of VMs of the unheated structure, i.e., \textit{cold modes}, in combination with some carefully selected \textit{dual} modes is found to be suitable for capturing the behavior of heated structures  \cite{Przekop_2007,Spottswood_2010}. It was further reported in ref.~\cite{Spottswood_2010} that the use of \emph{hot modes} in the basis leads to an improvement in results. In general, however, a practical strategy would be to construct a database of reduced-order bases for a set of static temperature fields and perform interpolation among these precomputed bases~\cite{ Amsallem, Amsallem_2009,Lieu_2004} to come up with ROMs for other values of temperature fields.

Now, if the temperature field is indeed static for a given structural dynamics problem, these approaches may be expected to perform well. However, when the temperature varies \emph{dynamically} during a structural simulation
(e.g., due to change in thermal boundary conditions or heat source), i.e.,
\begin{equation}
\label{eq:lin_example_time_varying}
M\ddot{x}+C\dot{x}+K(T(t))x=f(t)+g(T(t)),
\end{equation}
the aforementioned techniques are not directly applicable for reducing
structural dynamics equations due to two main reasons:
\begin{enumerate}
	\item \emph{Non-existence of an equilibrium point for local reduction}: As mentioned above, most reduction techniques usually construct relevant subspaces which are attached to an equilibrium point. However, in the case of dynamic temperature dependence, there is no such (unique) equilibrium. Even in the absence of mechanical loading, the internal force generated due to the dynamically varying temperature leads to a continuously ``changing'' equilibrium. Then, any local reduction technique would not be applicable since there is no unique equilibrium around which the dynamics can be approximated.  
	\item \emph{No invariant subspaces for reduction}: Each temperature configuration corresponds to a new set of VMs which form relevant  invariant subspaces for reduction, if temperature is kept static. However, due to a continuous change in temperature, we cannot find a fixed invariant subspace, even for the linear system \eqref{eq:lin_example_time_varying}. 
\end{enumerate}
Despite these issues, model reduction for structural equations has been successfully performed with dynamic temperature-dependence in some cases. One simple approach is to obtain a reference temperature configuration by averaging temperature variation over time for each point in space. This results in a static temperature field, around which a reduction basis can be calculated to reduce the problem where the temperature is changing dynamically. While this approach has been demonstrated to perform well in specific situations \cite{Falk_2011}, it is easy to envision scenarios where
this simple technique can lead to incorrect predictions. Indeed, consider
a beam under the influence of a spatial temperature pulse which is
oscillating along the length of the beam (modeling thermal effects
of an oscillating shock), as examined in ref.~\cite{Matney_2014}. In this case, the
above-mentioned averaging approach, would lead to a spatially
uniform temperature distribution, around which the reduction basis would be constructed. Clearly, such an approximation would be inconsistent with the physics of the problem. 

Furthermore, the above-mentioned technique of complimenting a basis of cold modes with dual modes would not be theoretically justifiable for reducing \eqref{eq:lin_example_time_varying} due to the two issues above. However, these techniques have been found to work well in specific cases \cite{Perez_2011,Matney_2014} where the structural dynamics is reduced under the influence of time-varying temperature distribution. 

In the context of dynamic parameter dependence, another technique~\cite{Tamarozzi_2014} involves the construction of a smooth parameter-dependent basis to reduce linear problems with application to moving external loads or boundary conditions, encountered during gear meshing. In this method, a time-dependent mapping from a low-dimensional space is introduced into the full system of unknowns such that 
\begin{equation}
\label{eq:time_dep_map}
u = V(p(t)) q(t)
\end{equation}  
where $ V $ is a basis matrix, parameterized with respect to the parameters $ p $, which are changing in time; and $ q(t) $ are the reduced set of unknowns. In this setting, a set of bases which are constructed for static parameter values are interpolated to obtain a family of bases $ V(p) $, which smoothly depends on the parameters $ p $. The time derivative of eq.~\eqref{eq:time_dep_map} leads to convective terms in the corresponding ROM, whose evaluation can pose a numerical challenge. It is important to note that unlike the modal subspaces which are invariant for static parameters, eq.~\eqref{eq:time_dep_map} describes an invariance relationship which does not hold in general, even for linear systems such as \eqref{eq:lin_example_time_varying}.
	\begin{figure}[H]
		\centering
		\begin{subfigure}{0.49\textwidth}
			\vspace{0.5cm}
			\centering\includegraphics[width=\linewidth]{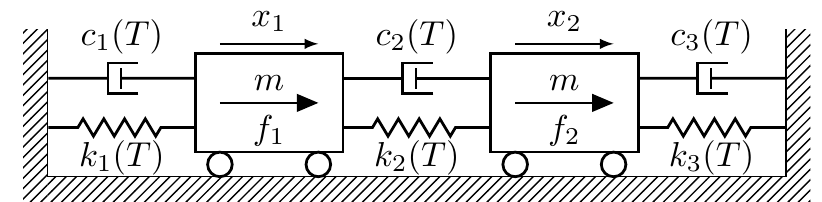}
			\vspace{1cm}
			\caption{\small A 2-DOF linear oscillator with $ T $-dependent properties}\label{fig:disconti_force}
		\end{subfigure}
		\begin{subfigure}{0.49\textwidth}
			\centering\includegraphics[width=0.7\linewidth]{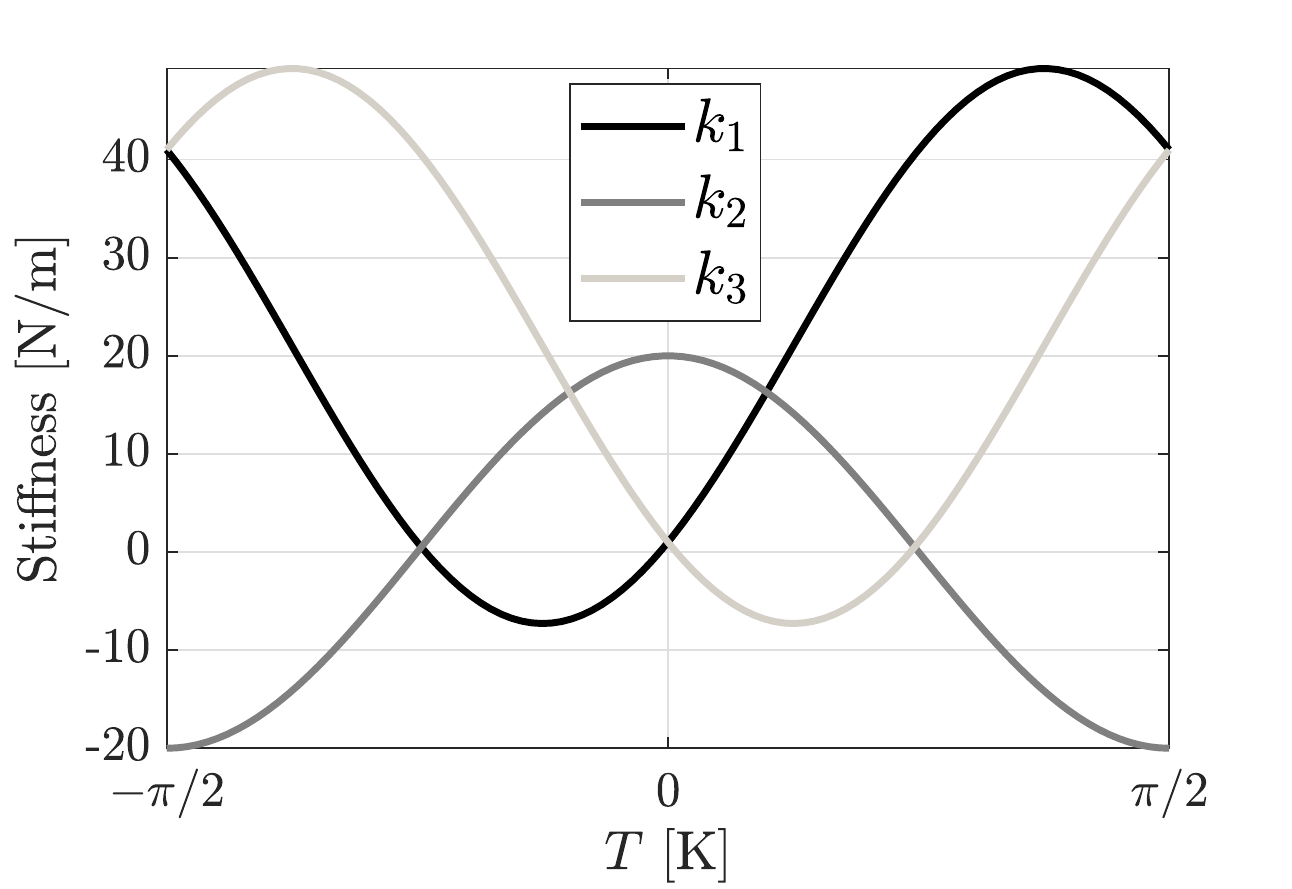}
			\caption{\small Variation in spring stiffness with respect to $ T $}\label{fig:disconti_NL}
		\end{subfigure}
		\begin{subfigure}{0.49\textwidth}
			\centering\includegraphics[width=0.9\linewidth]{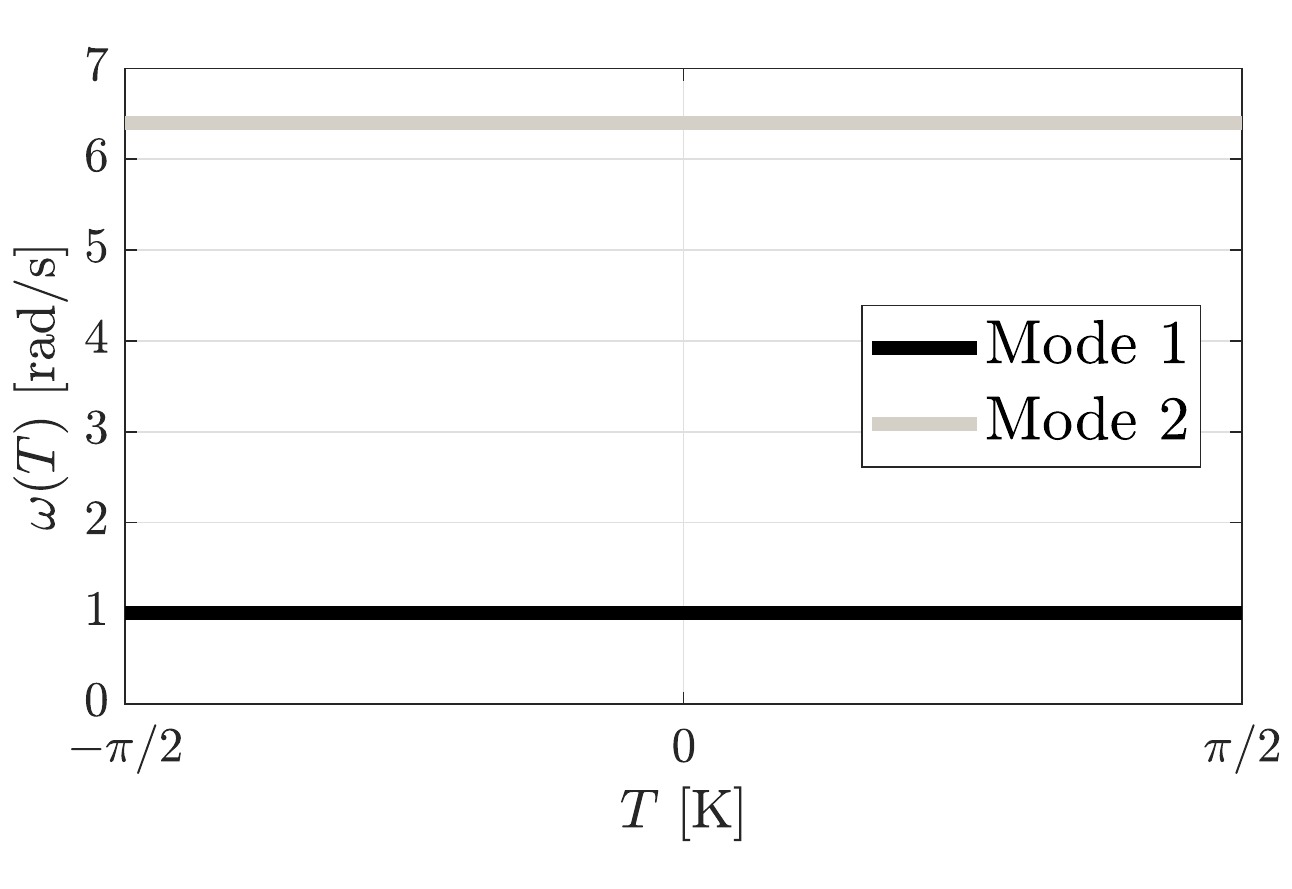}			
			\caption{\small Natural frequency ($ \omega $) of the two modes remain constant as a function of $ T $}
		\end{subfigure}
		\begin{subfigure}{0.49\textwidth}
			\centering\includegraphics[width=0.8\linewidth]{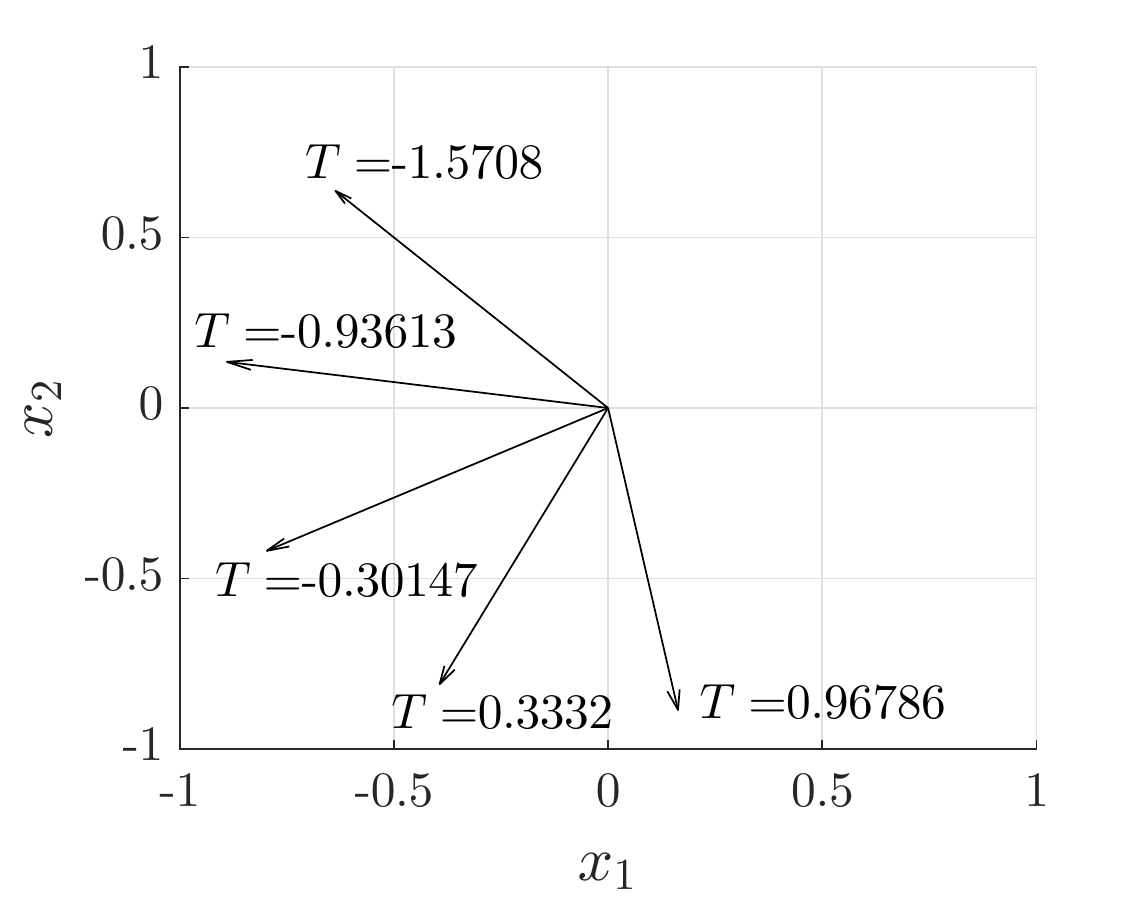}
			\caption{\small The first eigenmode at different values of $ T $}
		\end{subfigure}
		\caption{ \label{fig:2dof} \small (a) A \textbf{two degree-of-freedom oscillator} with identical masses $ m=1 $ Kg;  $ T \in [-\pi/2, \pi/2]$ represents the difference in system temperature from ambient temperature (K); $ k_1(T) = a + b [1+\cos(\alpha T)-\sin(\alpha T)],\, k_2(T) = b \cos(\alpha T),\, k_3(T) = a + b [1-\cos(\alpha T)-\sin(\alpha T)] $ are the temperature-dependent spring constants, where $ a = 1 $ N/m, $ b=20 $ N/m, $ \alpha =2 $ K$ ^{-1} $;  $ c_i(T) = \beta k_i(T) \,\forall i\in \{1,2,3\}$ are the proportional damping constants with $ \beta = 0.1 $ s. (b) Due to the specific choice of functions, the natural frequencies of the undamped system remain constant as function of $ T $. (c) The associated eigenvectors (VMs), however, change direction. (d) the spring constants as a function of temperature  }

	\end{figure}
   
\begin{figure}[H]
	\centering
	\begin{subfigure}{0.49\textwidth}
		\centering\includegraphics[width=\linewidth]{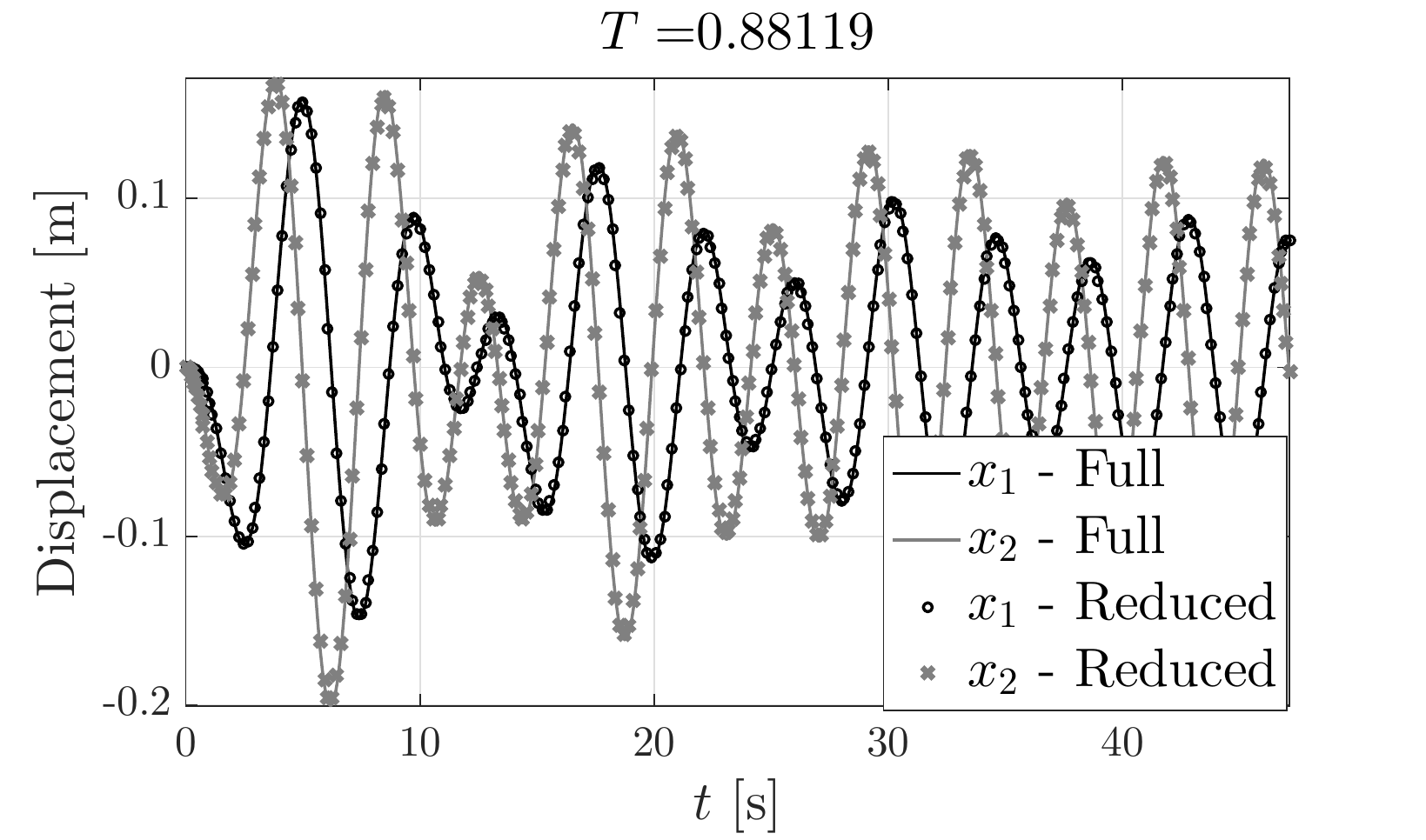}
		\caption{}
	\end{subfigure}
	\begin{subfigure}{0.49\textwidth}
		\centering\includegraphics[width=\linewidth]{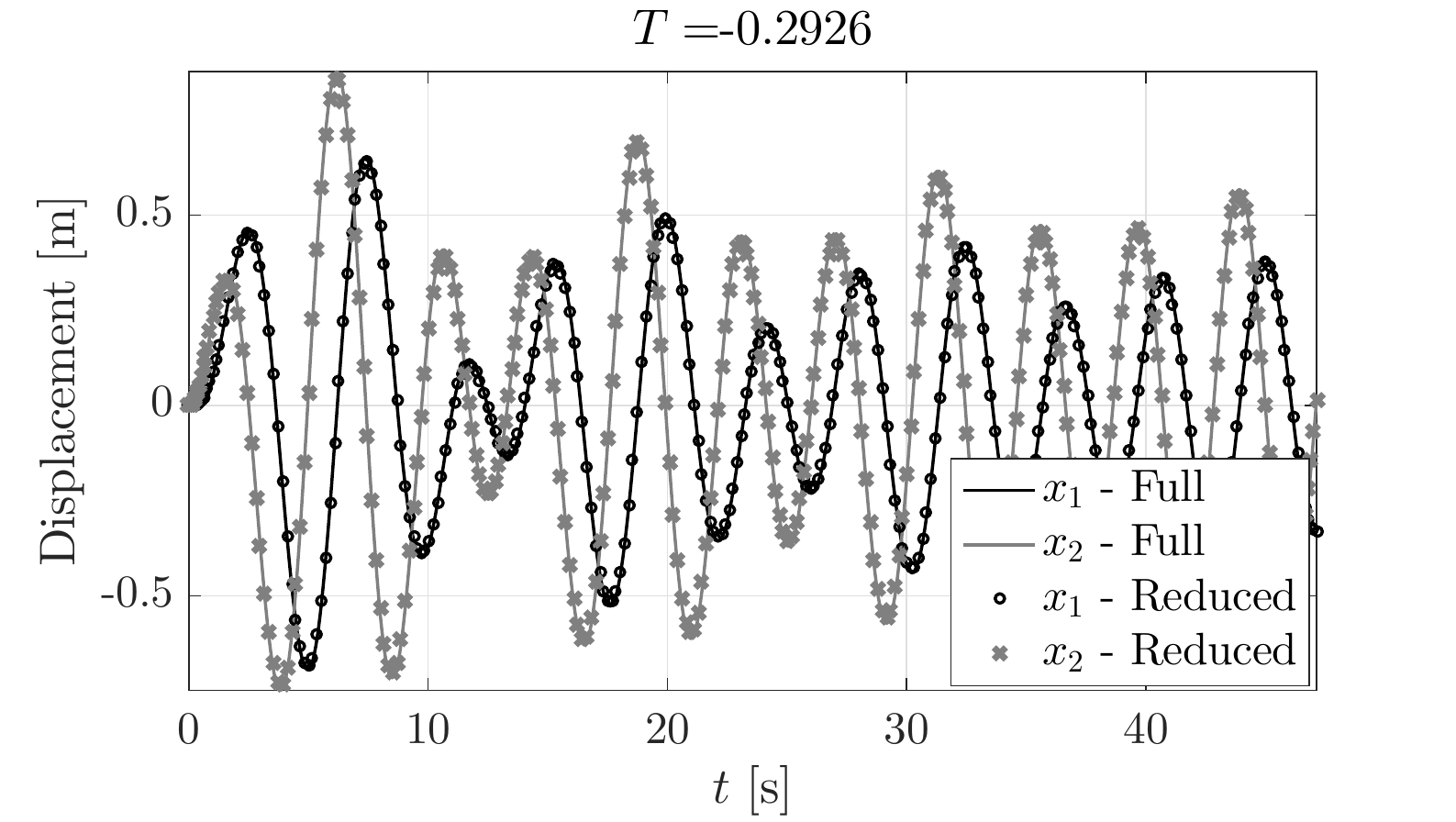}
		\caption{}
	\end{subfigure}
	\caption{ \small \label{fig:2dof_ROM_Tfixed} \small \textbf{Temperature constant}: Transient response of the two-degree-of-freedom oscillator (cf. Figure \ref{fig:2dof}) under the harmonic forcing $ [f_1(t), f_2(t)] = [0, \sin(\Omega t)] $ with forcing frequency $ \Omega = 1.5 $ rad/s for constant temperatures (a) $ T\approx 0.88$ K, (b) $ T\approx-0.29 $, with initial conditions $ x_1(0) = x_2(0) = 0 $. A single mode reduction using the first mode computed at the respective values of $ T $ is effective in model reduction. }
\end{figure}
Indeed, consider a simple two-degree-of-freedom linear mechanical oscillator, as shown in Figure~\ref{fig:2dof}. The spring and damping constant in this example are assumed to be temperature-dependent such that the eigen-frequencies of the system remain unchanged and well-separated for any variation in temperature. The eigenvectors, however, change upon varying temperature. While keeping the temperature constant, we force the system near its first natural frequency. This prompts us to reduce the system using the first eigenmode, resulting in a single-degree-of-freedom ROM at any fixed temperature. The same is depicted in Figure~\ref{fig:2dof_ROM_Tfixed} for a few of values of temperature.

Since the first mode calculated at different temperatures is found to capture the dynamics of the system in Figure \ref{fig:2dof}, the following questions arise in the case of dynamic temperature variation:
\begin{itemize}
	\item Is it reasonable that a single-mode basis which adapts instantaneously to the dynamic temperature change such as in eq.~\eqref{eq:time_dep_map} would be effective for model reduction?  
	\item And if yes, when can we expect such a reduction approach to produce good results? 
\end{itemize}
\begin{figure}[H]
	\centering
	\begin{subfigure}{0.49\textwidth}
		\centering\includegraphics[width=\linewidth]{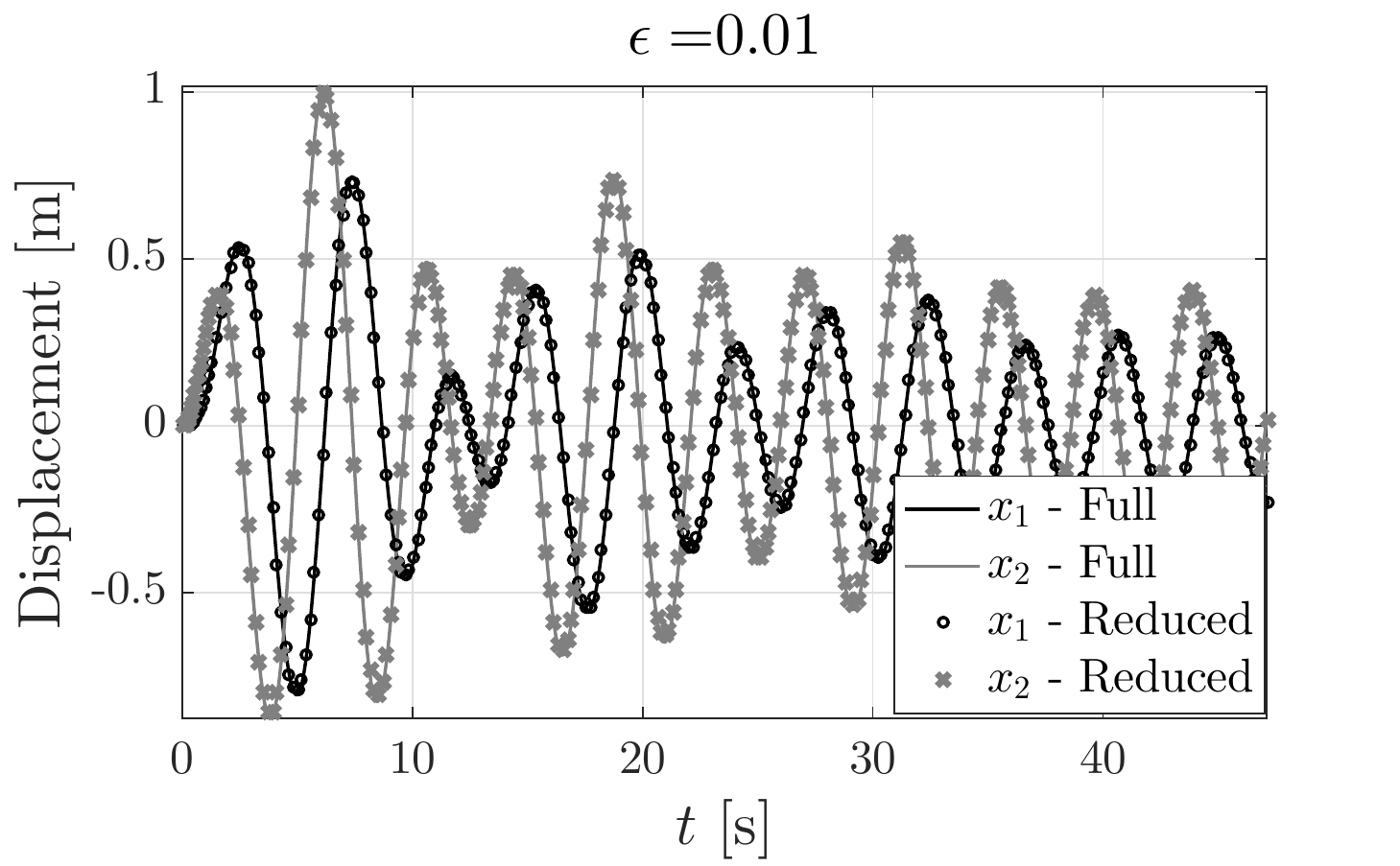}
		\caption{\small slow variation in $ T $}
	\end{subfigure}
	\begin{subfigure}{0.49\textwidth}
		\centering\includegraphics[width=\linewidth]{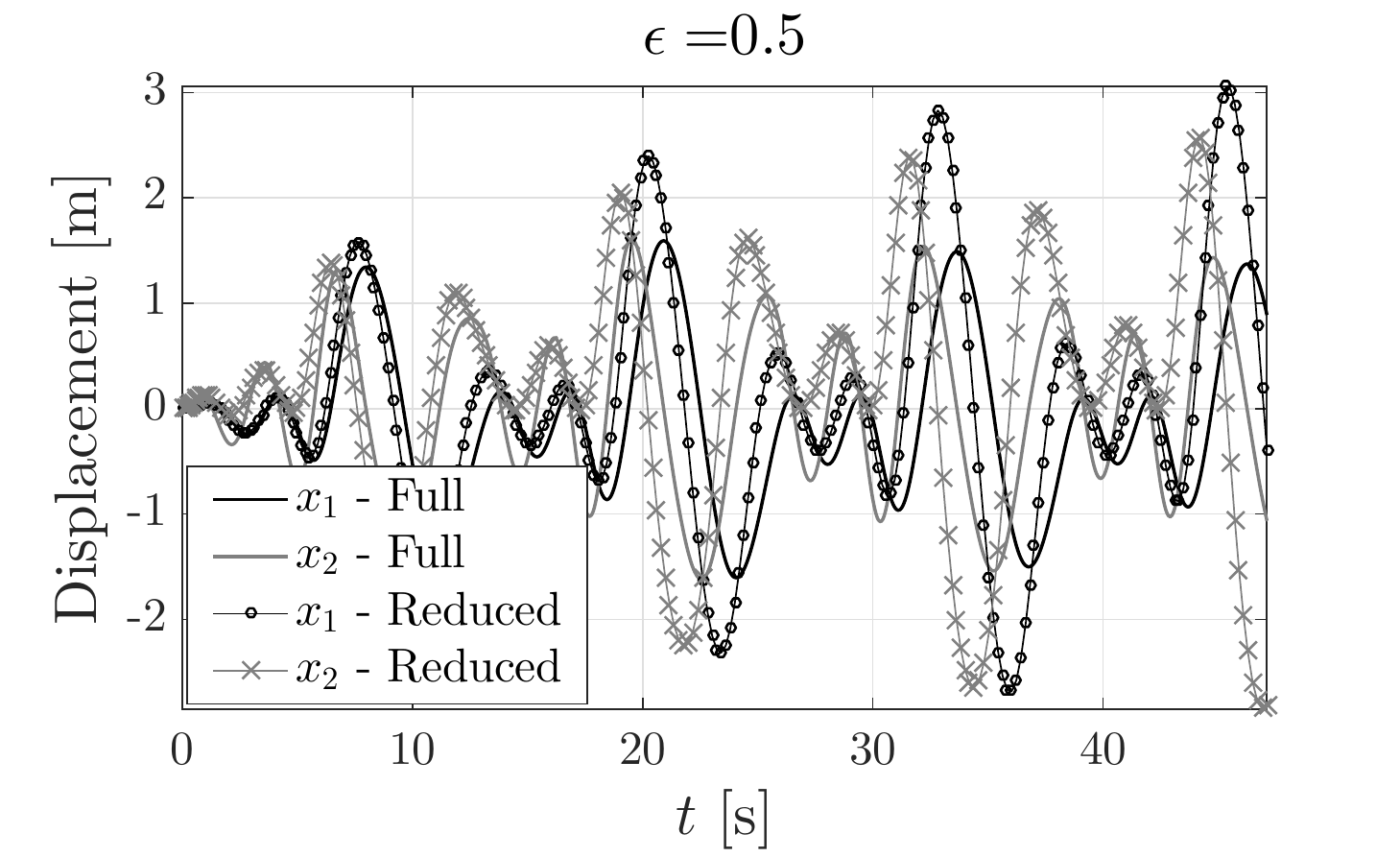}
		\caption{\small fast variation in $ T $}
	\end{subfigure}
	\caption{\small \label{fig:2dof_ROM_Tdyn} \small \textbf{Dynamic temperature variation}: Same as Figure \ref{fig:2dof_ROM_Tfixed} except $ T $ is dynamically varying as $ {T = (\pi/3) \sin(\epsilon t)} $ for: (a) $ \epsilon=0.01 $, i.e., when temperature is changing slowly, and (b) for $ \epsilon = 0.5 $, i.e., the temperature dynamics is evolving on a time-scale comparable to that of structural dynamics. (a) shows that the first VM, instantaneously adapting to temperature change  is effective for reduction (cf.~eq.~\eqref{eq:time_dep_map}) when $ T $ varies slowly in comparison with the system vibration frequency. However, as shown in~(b), this reduction fails when $ T $ varies relatively fast, even though the 2 modes have well-separated natural frequencies (cf.~Figure~\ref{fig:2dof}c) at all values of $ T $. }
\end{figure}

The answer to these question lies in an interesting observation, depicted in Figure \ref{fig:2dof_ROM_Tdyn}. We see that an instantaneously-adapting basis with a single mode is effective in capturing system dynamics under a slowly changing temperature. However, the same ROM is not effective when temperature varies rapidly.  This shows that a slow variation of parameters is necessary (but not sufficient) to guarantee effective model reduction using interpolation of bases obtained from static parameter values in the context of dynamic parameter dependence. 

To this end, we apply the method of multiple scales (MMS) (see, e.g., ref.~\cite{Kevorkian}) to the temperature-dependent structural dynamics equations, where the temperature is taken to be evolving at a different, slower time scale than that of the structural motion. This results in a set of partial differential equations (PDEs) in the slow thermal time and the fast structural time. Due to this treatment, the slowly varying temperature unknowns can be treated as static parameters in a mathematically justifiable manner. Interestingly, this procedure automatically avoids any convective terms arising from the time derivative of the mappings such as \eqref{eq:time_dep_map}, which are usually cumbersome to evaluate~\cite{Tamarozzi_2014}. 

Using the MMS, we obtain an asymptotic set of PDEs in the structural time scale, where the temperature can be seen as a slowly varying parameter. This allows for the calculation of a parameter-dependent equilibrium around which the aforementioned local reduction techniques can be applied using the corresponding subspaces in an adaptive manner. Thus, the use of MMS resolves both the issues mentioned above in this slow-fast, thermo-mechanical setting.

After applying the MMS to temperature-dependent structural equations in Section~\ref{sec:setup}, we first show the relevance of MMS on a simple single-degree-of-freedom thermo-mechanical oscillator. Next, we propose the use of a basis which instantaneously adapts to the slowly varying temperature in Section~\ref{sec:basis}. In doing, we preserve minimal number of unknowns in the basis due to changing temperature. In Section~\ref{sec:basis}, we first consider linear systems, reduced using temperature-dependent VMs, followed by geometrically nonlinear systems, reduced using VMs and modal derivatives in an adaptive basis. As a proof-of-concept, we test these propositions in Section~\ref{sec:results} on straight and curved beam examples in the linear as well as the geometrically nonlinear setting. Finally, some conclusion with scope for further work are laid down in Section~\ref{sec:conc}.  

\section{Multiple time scale dynamics}

\label{sec:setup}

After spatial discretization using finite elements (FE) of partial
differential equations (PDEs) governing temperature-dependent momentum
balance, we obtain a set of ordinary differential equations (ODEs).
Along with the initial conditions for generalized displacements and
velocities, these ODEs govern the temperature-dependent response of
the underlying structure in the form of an initial value problem as:
\begin{align}
\mathbf{M}\ddot{\mathbf{u}}+\mathbf{C}\dot{\mathbf{u}}+\mathbf{f}\left(\mathbf{u},\mathbf{T}\right) & =\mathbf{g}\left(t\right),\nonumber \\
\mathbf{u}(0) & =\mathbf{x}_{0},\label{eq:goveqn}\\
\dot{\mathbf{u}}(0) & =\mathbf{v}_{0},\nonumber 
\end{align}
where $\mathbf{u}(t)\in\mathbb{R}^{n}$ is the vector of generalized
displacements; $\mathbf{M}\in\mathbb{R}^{n\times n}$ is the mass
matrix; $\mathbf{C}\in\mathbb{R}^{n\times n}$ is the damping matrix;
$\mathbf{T}\in\mathbb{R}^{n_{T}}$ represents the spatially discretized
and possibly time-dependent temperature field prescribed on the
mechanical structure, $\mathbf{f}:\mathbb{R}^{n}\times\mathbb{R}^{n_{T}}\mapsto\mathbb{R}^{n}$
gives nonlinear elastic internal force as a function of the displacement
$\mathbf{u}$ and temperature $\mathbf{T}$ of the structure; and
$\mathbf{g}(t)\in\mathbb{R}^{n}$ is the time-dependent external load
vector, $\dot{(\bullet)}=\frac{d(\bullet)}{dt}$ denotes the derivative
with respect to time $t$. The system (\ref{eq:goveqn}) is sometimes
also referred to as the high fidelity model (HFM) in the literature.

As mentioned in the Introduction, the temperature field over structures
is expected to evolve much slower as compared to the structural time
scale $t$. The evolution of the temperature field over the structure
is usually modeled using the heat equation, which upon spatial discretization
(using, e.g., finite elements) gives 
\begin{equation}
\mathbf{M}_{T}\mathbf{T}^{\prime}+\mathbf{f}_{T}\left(\mathbf{T}\right)=\mathbf{h}(\tau),\label{eq:heat}
\end{equation}
where $\tau$ is the thermal time-scale; $(\bullet)^{\prime}$ denotes
the derivative with respect to $\tau$; $\mathbf{M}_{T}\in\mathbb{R}^{n_{T}\times n_{T}}$
is the thermal mass matrix; $\mathbf{f}_{T}:\mathbb{R}^{n_{T}}\mapsto\mathbb{R}^{n_{T}}$
gives the heat flux through the structure due to physical
effects such as conduction, convection, radiation, or a combination
there-of; $\mathbf{h}(\tau)$ represents the externally-applied heat
source to the structure. 

The slowness of the thermal dynamics with respect to the structural
one can be mathematically modeled with the relation $\tau=\epsilon t$,
where $0<\epsilon\ll1$ is a small physical parameter. In practice,
$\epsilon$ can be quantified as, e.g., the ratio of the time period
of the fundamental mode of oscillation of the structural system (\ref{eq:goveqn})
after linearization and the thermal time constant obtained from the
linearized spectral analysis of (\ref{eq:heat}). Now, if the temperature
field is a function of the slow time scale $\tau$, i.e., $\mathbf{T}=\mathbf{T}(\tau)$,
then we get 
\begin{equation}
\dot{\mathbf{T}}=\epsilon\mathbf{T}^{\prime}\,.\label{eq:Tdot}
\end{equation}
Equation (\ref{eq:Tdot}) suggests that if $\mathbf{T}$ naturally
evolves over the timescale $\tau$, i.e., $\mathbf{T}$ varies at
$\mathcal{O}(1)$ speed with respect to $\tau$, then it varies at
a much smaller $\mathcal{O}(\epsilon)$ speed over the structural
time scale $t$. Physically, the solution to the HFM~\eqref{eq:goveqn}
can be assumed to have a fast component due structural dynamics, as
well as a slow component under the influence of slowly changing temperature.
In other words, the solution evolves over two timescales as
\begin{equation}
\mathbf{u}=\mathbf{u}(t,\tau).
\end{equation}
The method of multiple scales (cf., e.g., ref.~\cite{Kevorkian}) then
allows us to express the solution in terms of an expansion in $\epsilon$
as 
\begin{equation}
\mathbf{u}(t,\tau)=\mathbf{u}_{0}(t,\tau)+\epsilon\mathbf{u}_{1}(t,\tau)+\epsilon^{2}\mathbf{u}_{2}(t,\tau)+\dots\,,\label{eq:exp}
\end{equation}
where $\mathbf{u}_{0},\mathbf{u}_{1},\mathbf{u}_{2}\dots$ are the
solution components at different orders in $\epsilon$, which can
be recursively obtained in the following manner. Using chain rule for the time derivative of eq.~\eqref{eq:exp}, we obtain
\begin{equation}
\begin{aligned}\dot{\mathbf{u}}= & \frac{\partial\mathbf{u}}{\partial t}+\frac{\partial\mathbf{u}}{\partial\tau}\frac{\partial\mathbf{\tau}}{\partial t}=\frac{\partial\mathbf{u}}{\partial t}+\epsilon\frac{\partial\mathbf{u}}{\partial\tau}\,,\\
\ddot{\mathbf{u}}= & \frac{\partial^{2}\mathbf{u}}{\partial t^{2}}+2\epsilon\frac{\partial^{2}\mathbf{u}}{\partial t\partial\tau}+\epsilon^{2}\frac{\partial^{2}\mathbf{u}}{\partial\tau^{2}}\,.
\end{aligned}
\label{eq:chainrule}
\end{equation}
Upon substituting \eqref{eq:exp} and \eqref{eq:chainrule} into the governing equations~\eqref{eq:goveqn}, we obtain a PDE in $t$ and $\tau$
given by
\begin{equation}
\mathbf{M}\left(\frac{\partial^{2}\mathbf{u}}{\partial t^{2}}+2\epsilon\frac{\partial^{2}\mathbf{u}}{\partial t\partial\tau}+\epsilon^{2}\frac{\partial^{2}\mathbf{u}}{\partial\tau^{2}}\right)+\mathbf{C}\left(\frac{\partial\mathbf{u}}{\partial t}+\epsilon\frac{\partial\mathbf{u}}{\partial\tau}\right)+\mathbf{f}\left(\mathbf{u}(t,\tau),\mathbf{T}(\tau)\right)=\mathbf{p}\left(t;\epsilon\right),\label{eq:pde}
\end{equation}
where we have introduced the dependence of the external mechanical
force $\mathbf{p}$ on $\epsilon$. We note that there is no reason for the external forcing to be dependent on the physical parameter $ \epsilon $. In this perturbative setting, however, we treat $ \epsilon $ as scaling constant that could be used to distinguish between components of a given forcing that have different magnitudes. Thus, $\mathbf{p}$ can be seen
as an appropriately-scaled version of $\mathbf{g}$ in \eqref{eq:goveqn},
e.g., if $\mathbf{g}$ is $ \mathcal{O}(\epsilon) $ small in comparison with other terms of the governing equations~\eqref{eq:goveqn}, then we can simply write
$\mathbf{g}(t)=\mathbf{p}(t,\epsilon)=\epsilon\mathbf{l}(t),$ where
$\mathbf{l}(t)$ is $\mathcal{O}(1)$. According to the MMS, eq.~\eqref{eq:pde} must be satisfied at all orders in
$\epsilon$. Using the expansion~\eqref{eq:exp} and after Taylor
expansion of the nonlinear internal force $\mathbf{f}$ around $\epsilon=0$,
we obtain the leading order terms in the PDE~\ref{eq:pde} as:
\begin{equation}
\text{\ensuremath{\mathcal{O}}}(1):\qquad\mathbf{M}\frac{\partial^{2}\mathbf{u}_{0}}{\partial t^{2}}+\mathbf{C}\frac{\partial\mathbf{u}_{0}}{\partial t}+\mathbf{f}\left(\mathbf{u}_{0},\mathbf{T}(\tau)\right)=\mathbf{p}\left(t,0\right).\label{eq:order1}
\end{equation}
The solution to eq.~\eqref{eq:order1} gives us the leading-order component~$\mathbf{u}_{0}$
in eq.~\eqref{eq:exp}. Equation~\eqref{eq:order1} can be integrated
in time $t$ using usual time integration schemes adopted for eq.~\eqref{eq:goveqn},
e.g., Newmark's scheme, with the same initial conditions, i.e., $\mathbf{u}_{0}(0)=\mathbf{x}_{0},\frac{\partial\mathbf{u}_{0}(0)}{\partial t}=\mathbf{v}_{0}$.
Collecting the $\mathcal{O}(\epsilon)$ terms in eq.~\eqref{eq:pde},
we obtain

\begin{align}
\text{\ensuremath{\mathcal{O}}}(\epsilon):\qquad\mathbf{M}\frac{\partial^{2}\mathbf{u}_{1}}{\partial t^{2}}+\mathbf{C}\frac{\partial\mathbf{u}_{1}}{\partial t}+\left[\frac{\partial\mathbf{f}}{\partial\mathbf{u}}\left(\mathbf{u}_{0},\mathbf{T}(\tau)\right)\right]\mathbf{u}_{1} & =-2\mathbf{M}\frac{\partial^{2}\mathbf{u}_{0}}{\partial t\partial\tau}-2\mathbf{C}\frac{\partial\mathbf{u}_{0}}{\partial\tau}+\frac{\partial\mathbf{p}}{\partial\epsilon}(t,0).\label{eq:ordereps}
\end{align}
Upon solving for $\mathbf{u}_{0}$ from eq.\eqref{eq:order1}, it is
easy to see that eq.~\eqref{eq:ordereps} represents a linear system of
equations in $\mathbf{u}_{1}$, the solution to which gives us an
$\mathcal{O}(\epsilon)$ correction to $\mathbf{u}_{0}$ as given
in eq.~\eqref{eq:exp}. Physically, \eqref{eq:ordereps} represents the
system linearized around the leading-order solution $\mathbf{u}_{0}$
with a forcing term on the right-hand-side (dependent on $\mathbf{u}_{0}$).
Since the initial conditions of the HFM~\eqref{eq:goveqn}) are independent
of $\epsilon$, they are included in the leading-order problem~\eqref{eq:order1},
shown above. Consequently, the initial conditions for solving eq.~\eqref{eq:ordereps}
should be identically zero, i.e., 
\[
\mathbf{u}_{1}(0)=\frac{\partial\mathbf{u}_{1}(0)}{\partial t}=\mathbf{0}\,.
\]

Thus, using the method of multiple scales, we obtain a series of problems
which\textendash as shall be shown later\textendash can be used to
systematically reduce the HFM at different orders in $\epsilon$.
Equations~\eqref{eq:order1} and \eqref{eq:ordereps} give only the
equations for $\mathbf{u}_{0}$ and $\mathbf{u}_{1}$. Higher-order
terms in eq.~\eqref{eq:exp} can be found by following a similar recursive
procedure by collecting higher-order terms in eq.~\eqref{eq:pde}. Furthermore,
as is the case with the $\mathcal{O}(\epsilon)$ problem~\eqref{eq:ordereps},
it easy to see that all higher problems would be linear in their corresponding
unknowns. 

\section{Adaptive basis for model reduction}

\label{sec:basis}

\subsection{Linear systems: reduction using vibration modes}
For the sake of consistency, we first briefly review the concept of modal super-position which is the classic choice for modal reduction of linear mechanical system. 

\subsubsection{Static temperature dependence}
It is well known that modal subspaces (i.e., subspaces spanned by
VMs) are invariant and useful in reduction of linear systems.
This is also referred to as modal superposition in literature (cf.,
e.g., G\'{e}radin \& Rixen~\cite{Rixen97}). Consider the case of linear structural dynamics
independent of temperature or when the temperature configuration remains fixed during a dynamic simulation, i.e., $\mathbf{f}(\mathbf{u},\mathbf{T})=\mathbf{K}(\mathbf{T}_0)\mathbf{u}+\mathbf{b}(\mathbf{T_0})$
in eq.~\eqref{eq:goveqn} such that 
\begin{equation}
\mathbf{M}\ddot{\mathbf{u}}+\mathbf{C}\dot{\mathbf{u}}+\mathbf{K}(\mathbf{T}_0)\mathbf{u}+\mathbf{b}(\mathbf{T_0})=\mathbf{g}\left(t\right),\label{eq:lin}
\end{equation}
where $ \mathbf{T}_0$ is a fixed temperature configuration. In the absence of dynamic excitation $\mathbf{g}(t)$, the above system
has a unique equilibrium given as 
\begin{align*}
\mathbf{u}_{eq} & =-\mathbf{K}(\mathbf{T}_0)^{-1}\mathbf{b}(\mathbf{T}_0).
\end{align*}
Upon shifting the origin to
$\mathbf{u}_{eq}$, we get 
\begin{equation}
\mathbf{M}\ddot{\mathbf{y}}+\mathbf{C}\dot{\mathbf{y}}+\mathbf{K}(\mathbf{T}_0)\mathbf{y}=\mathbf{g}\left(t\right),\label{eq:linsys}
\end{equation}
where $ \mathbf{y} = \mathbf{u} - \mathbf{u}_{eq} $. The
VMs of system~\eqref{eq:linsys} form a basis for the generalized displacements~$ \mathbf{y} $ around the equilibrium $ \mathbf{u}_{eq} $. 
For proportionally damped structures, the undamped VMs $ \boldsymbol{\phi}_{i} $  that are obtained from the solution of the eigenvalue problem, 
\begin{equation}
\left(\mathbf{K}(\mathbf{T}_0)-\omega_{i}^{2}\mathbf{M}\right)\boldsymbol{\phi}_{i}=\mathbf{0}, \quad i=1,\dots,n
\end{equation}
lead to decoupling of the system~\eqref{eq:linsys} and are
used in a modal truncation approximation as
\begin{equation}
\mathbf{y}(t)=\sum_{i=1}^{n}\boldsymbol{\phi}_{i}q_{i}(t)\approx\sum_{i\in M}\boldsymbol{\phi}_{i}q_{i}(t),
\end{equation}
where $\mathcal{M}\subset\{1,2,\dots,n\}$ is a set containing the indices of
the VMs relevant for modal truncation (cf. G\'{e}radin \& Rixen ~\cite{Rixen97}), such that $m:=|\mathcal{M}|\ll n$. \\
The corresponding ROM is obtained via the classic Galerkin projection as 
\begin{equation}
	\mathbf{\Phi}^{\top}\mathbf{M}\mathbf{\Phi}\ddot{\mathbf{q}}+\mathbf{\Phi}^{\top}\mathbf{C}\mathbf{\Phi}\dot{\mathbf{q}}+\mathbf{\Phi}^{\top}\mathbf{K}\mathbf{\Phi}\mathbf{q}=\mathbf{\Phi}^{\top}\mathbf{g}\left(t\right),\label{eq:linsysred}
\end{equation}
where $\boldsymbol{\Phi}\in\mathbb{R}^{n\times m}$ is a matrix containing
the VMs indexed by the set $\mathcal{M}$. The solution
to the original system~\eqref{eq:lin} can then be recovered as 
\begin{equation}
\mathbf{\implies u}(t)\approx\mathbf{u}_{eq}+\sum_{i\in M}\boldsymbol{\phi}_{i}q_{i}(t)=\mathbf{u}_{eq}+\boldsymbol{\Phi}\mathbf{q}(t)\,.
\end{equation}

\subsubsection{Presence of dynamic temperature-dependence}

If we allow temperature dependence in an arbitrary fashion, such that
\begin{equation}
\mathbf{M}\ddot{\mathbf{u}}+\mathbf{C}\dot{\mathbf{u}}+\mathbf{K}\left(\mathbf{T}(t)\right)\mathbf{u}+\mathbf{b}(\mathbf{T}(t))=\mathbf{g}\left(t\right),\label{eq:lin_temp}
\end{equation}
we see that even in the absence of external forcing $ \mathbf{g}(t) $, the system \eqref{eq:lin_temp} has no equilibrium point due to the presence of dynamic thermal excitation $\mathbf{b}(\mathbf{T}(t))$. Thus, due to the absence of equilibrium and dynamic temperature variation, one cannot use specific VMs to reduce the system, even in a linear system such as \eqref{eq:lin_temp}, as we discussed in the Introduction. 

This problem is commonly tackled in the literature (cf. refs.~\cite{Przekop_2007,Spottswood_2010,Falk_2011,Perez_2011,Matney_2014}) by simply treating the temperature-dependent term $ \mathbf{b}(\mathbf{T}(t))$  as yet another time-dependent forcing term similar to $ \mathbf{g}(t) $. Specifically, we obtain the system 
	\begin{equation}
	\mathbf{M}\ddot{\mathbf{u}}+\mathbf{C}\dot{\mathbf{u}}+\mathbf{K}\left(\mathbf{T}(t)\right)\mathbf{u} = \tilde{\mathbf{g}}\left(t\right), \label{eq:lin_temp2}
	\end{equation}
	where $ \tilde{\mathbf{g}}\left(t\right):= {\mathbf{g}}\left(t\right)-\mathbf{b}(\mathbf{T}(t)) $. Note that the linear system~\eqref{eq:lin_temp2} now exhibits a unique trivial equilibrium point, i.e., $ \mathbf{u}_{eq}=\mathbf{0} $ in the absence of forcing $ \tilde{\mathbf{g}}\left(t\right) $. This manipulation leads to the routinely performed Galerkin projection about the origin via the mapping
	\begin{equation}
		\mathbf{u}(t) \approx\mathbf{V}\mathbf{q}(t).
	\end{equation}
	The reduction basis $ \mathbf{V} $ is usually constructed in the literature by an ad hoc selection of modes (hot/cold/dual modes) that are expected to account for the mechanical forcing as well as the dynamic temperature dependence, as discussed in the Introduction.

However, if it is a priori known
that the thermal dynamics is \emph{slow}, as is usually the case for
thermo-mechanical systems, we can apply the method of multiple scales,
as described in Section~\ref{sec:setup}, to obtain the leading order
problems as follows (cf.~eq.~\eqref{eq:order1}):
\begin{equation}
\mathbf{M}\frac{\partial^{2}\mathbf{u}_{0}}{\partial t^{2}}+\mathbf{C}\frac{\partial\mathbf{u}_{0}}{\partial t}+\mathbf{K}\left(\mathbf{T}(\tau)\right)\mathbf{u}+\mathbf{b}(\mathbf{T}(\tau))=\mathbf{p}\left(t,0\right).\label{eq:order1linear}
\end{equation}
In this PDE in time $t$, the slow time $\tau$ can be seen as parameter,
dependent upon which equilibria can be calculated for the unforced
systems as 
\begin{equation}
\mathbf{u}_{eq}(\tau)=-\mathbf{K}\left(\mathbf{T}(\tau)\right)^{-1}\mathbf{b}(\mathbf{T}(\tau)).\label{eq:eqmlin}
\end{equation}
Furthermore, the solution to the eigenvalue problem
\begin{equation}
\left[\mathbf{K}(\mathbf{T}(\tau))-\omega_{i}^{2}(\tau)\mathbf{M}\right]\boldsymbol{\phi}_{i}(\tau)=\mathbf{0}
\end{equation}
results in the temperature-dependent VMs $ \boldsymbol{\phi}_{i}(\tau) $ around this manifold of equilibria. The solution to the leading order problem~\eqref{eq:order1linear} is then approximated using a parameterized basis~$\mathbf{\boldsymbol{\Phi}}(\tau)$ containing a truncated set of VMs as 
\begin{equation}\label{eq:leading_order_linear}
\mathbf{u}_{0}(t,\tau)\approx\mathbf{u}_{eq}(\tau)+\boldsymbol{\Phi}(\tau)\mathbf{q}_{0}(t).
\end{equation}
A reduced-order model can be obtained by projecting the leading order
system~\eqref{eq:order1linear} on to the slowly varying basis as 
\begin{align*}
\boldsymbol{\Phi}(\tau)^{\top}\left[\mathbf{M}\boldsymbol{\Phi}(\tau)\frac{\partial^{2}\mathbf{q}_{0}}{\partial t^{2}}+\mathbf{C}\boldsymbol{\Phi}(\tau)\frac{\partial\mathbf{q}_{0}}{\partial t}+\mathbf{K}\left(\mathbf{T}(\tau)\right)\left[\mathbf{u}_{eq}(\tau)+\boldsymbol{\Phi}(\tau)\mathbf{q}_{0}(t)\right]+\mathbf{b}(\mathbf{T}(\tau))\right] & =\boldsymbol{\Phi}(\tau)^{\top}\mathbf{p}\left(t,0\right)\,
\end{align*}
and using eq.~\eqref{eq:eqmlin} we obtain the following reduced order model at the leading order
\begin{equation}
\small
\label{eq:linO1}
\mathcal{O}(1):\qquad\underbrace{\boldsymbol{\Phi}(\tau)^{\top}\mathbf{M}\boldsymbol{\Phi}(\tau)}_{\mathbf{M}_{\boldsymbol{\Phi}}(\tau)}\frac{\partial^{2}\mathbf{q}_{0}}{\partial t^{2}}+\underbrace{\boldsymbol{\Phi}(\tau)^{\top}\mathbf{C}\boldsymbol{\Phi}(\tau)}_{\mathbf{C}_{\boldsymbol{\Phi}}(\tau)}\frac{\partial\mathbf{q}_{0}}{\partial t}+\underbrace{\boldsymbol{\Phi}(\tau)^{\top}\mathbf{K}\left(\mathbf{T}(\tau)\right)\boldsymbol{\Phi}(\tau)}_{\mathbf{K}_{\boldsymbol{\Phi}}(\tau)}\mathbf{q}_{0}(t) =\boldsymbol{\Phi}(\tau)^{\top}\mathbf{p}\left(t,0\right)\quad
\end{equation}
Furthermore, the $\mathcal{O}(\epsilon)$ correction according to
eq.~\eqref{eq:order1} is given by 
\begin{align}
\mathbf{M}\frac{\partial^{2}\mathbf{u}_{1}}{\partial t^{2}}+\mathbf{C}\frac{\partial\mathbf{u}_{1}}{\partial t}+\mathbf{K}\left(\mathbf{T}(t)\right)\mathbf{u}_{1} =-2\mathbf{M}\frac{\partial^{2}\mathbf{u}_{0}}{\partial t\partial\tau}-2\mathbf{C}\frac{\partial\mathbf{u}_{0}}{\partial\tau}+\frac{\partial\mathbf{p}}{\partial\epsilon}(t,0).\label{eq:ordereps-1-1-1}
\end{align}
Using the same modal basis to reduce this linear problem and observing
that origin is the unique fixed point for this system, we get 
\begin{equation}
\mathbf{u}_{1}(t,\tau)\approx\boldsymbol{\Phi}(\tau)\mathbf{q}_{1}(t)\,,
\end{equation}
which upon Galerkin projection yields the reduced-order model at $ \mathcal{O}(\epsilon) $
\begin{align}
\mathcal{O}(\epsilon): \qquad \mathbf{M}_{\boldsymbol{\Phi}}\frac{\partial^{2}\mathbf{q}_{1}}{\partial t^{2}}+\mathbf{C}_{\boldsymbol{\Phi}}\frac{\partial\mathbf{q}_{1}}{\partial t}+\mathbf{K}_{\boldsymbol{\Phi}}(\tau)\mathbf{q}_{1} &= \boldsymbol{\Phi}^{\top}\left[\frac{\partial\mathbf{p}}{\partial\epsilon}(t,0)-2\mathbf{M}\frac{\partial^{2}\mathbf{u}_{0}}{\partial t\partial\tau}-2\mathbf{C}\frac{\partial\mathbf{u}_{0}}{\partial\tau}\right]\\
&= \boldsymbol{\Phi}^{\top}\left[\frac{\partial\mathbf{p}}{\partial\epsilon}(t,0)-2\mathbf{M}\mathbf{\Phi}^{\prime} \frac{\partial\mathbf{q}_{0}}{\partial t}-2\mathbf{C}\mathbf{\Phi}^{\prime}\mathbf{q}_0\right]\,.
\end{align}
Thus, in the linear case, the reduced operators are the same for reduced
problems at the leading order, as well as at $\mathcal{\mathcal{O}}(\epsilon)$.
In fact, it is easy to see that these linear operators remain the
same for higher order corrections, as well. 

We would like to emphasize that although this formal procedure can
be carried out even if the underlying thermal dynamics is not slow,
i.e., cases where $\epsilon$ is not small enough, it would not be
mathematically justifiable and can very easily lead to spurious results, as shown in the example in Section~\ref{sec:introduction}.

\subsection{Geometrically nonlinear systems}
\label{sec:MMS_nl}
For reduction of linear systems, a basis comprising of a truncated set of VMs can be easily identified by the examination of the external forcing and the linear spectrum. However, such a basis is usually not suitable for reduction of geometrically nonlinear systems because of membrane effects, which become significant in the nonlinear regime due to the bending-stretching-torsion coupling. In such cases, modal derivatives have emerged as a simple and effective model-reduction tool (cf.~refs.~\cite{IC,IC2,qm,qm2,WWS}). We discuss the use of modal derivatives in the context of dynamic temperature dependence.

The temperature-dependent equilibrium $\mathbf{u}_{eq}\left(\tau\right)$ is given by the solution to static problem
\begin{equation}
\mathbf{f}\left(\mathbf{u}_{eq}\left(\tau\right),\mathbf{T}(\tau)\right)=\mathbf{0}\,. 
\end{equation}
We linearize the nonlinear internal force $ \mathbf{f}$ around the configuration $ \mathbf{u}_{eq}\left(\tau\right) $ to obtain
the tangent stiffness matrix $\frac{\partial\mathbf{f}}{\partial\mathbf{u}}\left(\mathbf{u}_{eq}(\tau),\mathbf{T}(\tau)\right)$ evaluated at the temperature-dependent
equilibrium $ \mathbf{u}_{eq}\left(\tau\right) $. Since $ \tau $ is taken as a slowly varying parameter, we can compute the temperature-dependent VMs of the structure similar to the linear case as
\begin{equation}
\label{eq:nlin_VMs}
\left[\frac{\partial\mathbf{f}}{\partial\mathbf{u}}\left(\mathbf{u}_{eq}(\tau),\mathbf{T}(\tau)\right)-\omega_{i}^{2}(\tau)\mathbf{M}\right]\boldsymbol{\phi}_{i}(\tau)=\mathbf{0}.
\end{equation}
The static modal derivatives (cf.~ref.~\cite{qm}), which are relevant for capturing the membrane effects in thin-walled structures are given as 
\begin{equation}
\label{eq:MDs}
\mathbf{K}_{t}\left(\mathbf{u}_{eq}(\tau),\mathbf{T}(\tau)\right)\boldsymbol{\theta}_{ij}=-\left[\frac{\partial^{2}\mathbf{f}}{\partial\mathbf{u}\partial\mathbf{u}}\left(\mathbf{u}_{eq},\mathbf{T}(\tau)\right)\cdot\boldsymbol{\phi}_{j}\right]\boldsymbol{\phi}_{i},
\end{equation}
where $ \boldsymbol{\theta}_{ij} $ is the static modal derivative and physically represents the change in mode $ \boldsymbol{\phi}_i $ as the structure is perturbed in the direction of mode $ \boldsymbol{\phi}_j $.  Note that here the modal derivatives are calculated around a non-trivial, temperature-dependent equilibrium. This is in contrast with the case of no temperature dependence usually treated in literature, where the modal derivatives are simply constructed around the origin.

The reduction basis is formed by combining a truncated set of modes with the corresponding modal derivatives. As observed in ref.~\cite{qm}, a set of $ k $ VMs results in $ k(k+1)/2 $ modal derivatives. These modal derivatives may be selected to further reduce the basis size (cf. refs.~\cite{MasterThesis,Tiso}).
\begin{equation}
\mathbf{V}(\tau)=[\boldsymbol{\Phi}(\tau),\quad\boldsymbol{\Theta}(\tau)],
\end{equation}
where $ \mathbf{V}(\tau) \in \mathbb{R}^{n\times m}$ is the reduction basis.
Analogous to the linear case (cf. \eqref{eq:leading_order_linear}), we express the leading-order reduced solution as  
\begin{equation}
\label{eq:leading_order_nl}
\mathbf{u}_{0}(t,\tau)\approx\mathbf{u}_{eq}(\tau)+\mathbf{V}(\tau)\mathbf{q}_{0}(t)
\end{equation}
Upon substituting \eqref{eq:leading_order_nl} into the leading order MMS expansion \eqref{eq:order1} and performing Galerkin projection, we get
\begin{equation}
\mathbf{V}(\tau)^{\top}\mathbf{M\mathbf{V}(\tau)}\frac{\partial^{2}\mathbf{q}_{0}}{\partial t^{2}}+\mathbf{V}(\tau)^{\top}\mathbf{C}\mathbf{V}(\tau)\frac{\partial\mathbf{q}_{0}}{\partial t}+\mathbf{V}(\tau)^{\top}\mathbf{f}\left(\mathbf{u}_{eq}(\tau)+\mathbf{V}(\tau)\mathbf{q}_{0}(t),\mathbf{T}(\tau)\right)=\mathbf{V}(\tau)^{\top}\mathbf{p}\left(t,0\right)\,.\label{eq:Order1red}
\end{equation}
The $ \mathcal{O}(\epsilon) $ correction $ \mathbf{u}_1 $ given in \eqref{eq:ordereps} can be reduced using the same adaptive basis as
\begin{equation}
\mathbf{u}_{1}(t,\tau)\approx\mathbf{V}(\tau)\mathbf{q}_{1}(t).
\end{equation}
And again, by means of Galerkin projection, we obtain a reduced version of \eqref{eq:ordereps} which is linear in the reduced unknowns $ \mathbf{q}_1 $ as
\begin{equation}
\small
\mathbf{M}_{\mathbf{V}}(\tau)\frac{\partial^{2}\mathbf{q}_{1}}{\partial t^{2}}+\mathbf{C}_{\mathbf{V}}(\tau)\frac{\partial\mathbf{q}_{1}}{\partial t}+\left[\mathbf{V}(\tau)^{\top}\left(\frac{\partial\mathbf{f}}{\partial\mathbf{u}}\left(\mathbf{u}_{0},\mathbf{T}(\tau)\right)\right)\mathbf{V}(\tau)\right]\mathbf{q}_{1}=\mathbf{V}(\tau)^{\top}\left[\frac{\partial\mathbf{p}}{\partial\epsilon}(t,0)-2\mathbf{M}\frac{\partial^{2}\mathbf{u}_{0}}{\partial t\partial\tau}-2\mathbf{C}\frac{\partial\mathbf{u}_{0}}{\partial\tau}\right].\label{eq:orderepsred}
\end{equation}

In the special case of a small $\mathcal{O}(\epsilon)$ mechanical forcing,
e.g. $\mathbf{p}(t,\epsilon)=\mathbf\epsilon\mathbf{l}(t)$
we have
\begin{equation}
\label{eq:small_forcing}
\mathbf{M}\ddot{\mathbf{u}}+\mathbf{C}\dot{\mathbf{u}}+\mathbf{f}\left(\mathbf{u},\mathbf{T}\right)=\epsilon\mathbf{l}(t).
\end{equation}
and the leading-order ROM is given by 
\begin{equation}
\mathcal{O}(1):\qquad\mathbf{M}_{\mathbf{V}}(\tau)\frac{\partial^{2}\mathbf{q}_{0}}{\partial t^{2}}+\mathbf{C}_{\mathbf{V}}(\tau)\frac{\partial\mathbf{q}_{0}}{\partial t}+\mathbf{V}(\tau)^{\top}\mathbf{f}\left(\mathbf{u}_{eq}(\tau)+\mathbf{V}(\tau)\mathbf{q}_{0}(t),\mathbf{T}(\tau)\right)=\mathbf{0},
\end{equation}
which is the system response to thermal loading only. This would be
a good approximation of the system response for small enough $\epsilon$
as long $\mathbf{u}_{eq}$ is a stable equilibrium. Indeed, the mechanical
system would be performing small-amplitude oscillations around the
applied thermal loading. These can be captured by the $\mathcal{O}(\epsilon)$
ROM as 
\begin{equation}
\small
\mathbf{M}_{\mathbf{V}}(\tau)\frac{\partial^{2}\mathbf{q}_{1}}{\partial t^{2}}+\mathbf{C}_{\mathbf{V}}(\tau)\frac{\partial\mathbf{q}_{1}}{\partial t}+\left[\mathbf{V}(\tau)^{\top}\left(\frac{\partial\mathbf{f}}{\partial\mathbf{u}}\left(\mathbf{u}_{0},\mathbf{T}(\tau)\right)\right)\mathbf{V}(\tau)\right]\mathbf{q}_{1}=\mathbf{V}(\tau)^{\top}\left[\mathbf{l}(t)-2\mathbf{M}\frac{\partial^{2}\mathbf{u}_{0}}{\partial t\partial\tau}-2\mathbf{C}\frac{\partial\mathbf{u}_{0}}{\partial\tau}\right]\,.
\end{equation}
\subsection{Basis interpolation and mode veering}

Though the chosen subspace spanned by basis vectors at each temperature
configuration would be uniquely defined, the same is not true for
the basis representation of such a subspace. Each subspace would be
defined by an equivalence class of orthogonal bases. There is a possibility
of mode veering and rotation between bases calculated at two different
temperature configurations if the representative basis for a given
temperature configuration is chosen arbitrarily. This would pose a
serious issue in adaptive model reduction, as proposed in this work.
Indeed, the modal amplitudes in ROM equations~\eqref{eq:Order1red}),\eqref{eq:orderepsred} must change continuously in time. This is
generally not true if the corresponding basis vectors are changing
discontinuously, as would be expected when mode veering and/or rotation
occurs between bases. A method to avoid this issue was proposed in ref.~\cite{Amsallem} by identifying a congruence transformation which transforms a basis such that its
vectors are consistent with those in a selected reference basis. We reproduce the procedure from ref.~\cite{Amsallem} in Algorithm~\ref{alg:congruent_basis}. 

\begin{algorithm}[H]
\begin{algorithmic}[1]

\Require{Reference basis $ \mathbf{V}_0 $, other bases $\tilde{\mathbf{V}}_1, \tilde{\mathbf{V}}_2, \dots, \tilde{\mathbf{V}}_N $}

\Ensure{Bases ${\mathbf{V}}_1, {\mathbf{V}}_2, \dots, {\mathbf{V}}_N $, congruent to $ \mathbf{V}_0 $}	

\Statex

\For{$j \gets 1 \textrm{ to } N$}

\State $\mathbf{P}_{j}=\tilde{\mathbf{V}}_{j}^{\top}\mathbf{V}_{0}$

\State $[\mathbf{L}_{j},\boldsymbol{\Sigma}_{j},\mathbf{R}_{j}]=\texttt{svd}(\mathbf{P}_{j})$
\Comment Singular value decomposition (SVD) of $\mathbf{P}_{j}$,
i.e., $\mathbf{P}_{j}=\mathbf{L}_{j}\boldsymbol{\Sigma}_{j}\mathbf{R}_{j}^{\top}$

\State $\mathbf{Q}_{j}=\mathbf{L}_{j}\mathbf{R}_{j}^{\top}$

\State $\mathbf{V}_{j}=\tilde{\mathbf{V}}_{j}\mathbf{Q}_{j}$

\EndFor

\end{algorithmic}

\caption{\label{alg:congruent_basis}Setting up congruent set of bases}
\end{algorithm}

\section{Numerical examples}
\label{sec:results}
We demonstrate the proposed method on a set of beam examples, with displacements in 2-dimensional $(x,z)$ plane. The beam is modeled using finite elements with a linear elastic material (see Table~\ref{tab:props} for geometrical parameters and physical properties). Furthermore, we consider geometric nonlinearities in the model via the von-K\'{a}rm\'{a}n strain approximation. The thermal effects are modelled by assuming that the temperature distribution is uniform throughout the thickness of the beam. Across all examples, the beam is clamped at both ends and discretized using 60 elements. 
\begin{figure}[H]
	\centering
	\begin{subfigure}{0.48\textwidth}
		\centering\includegraphics[width=\linewidth]{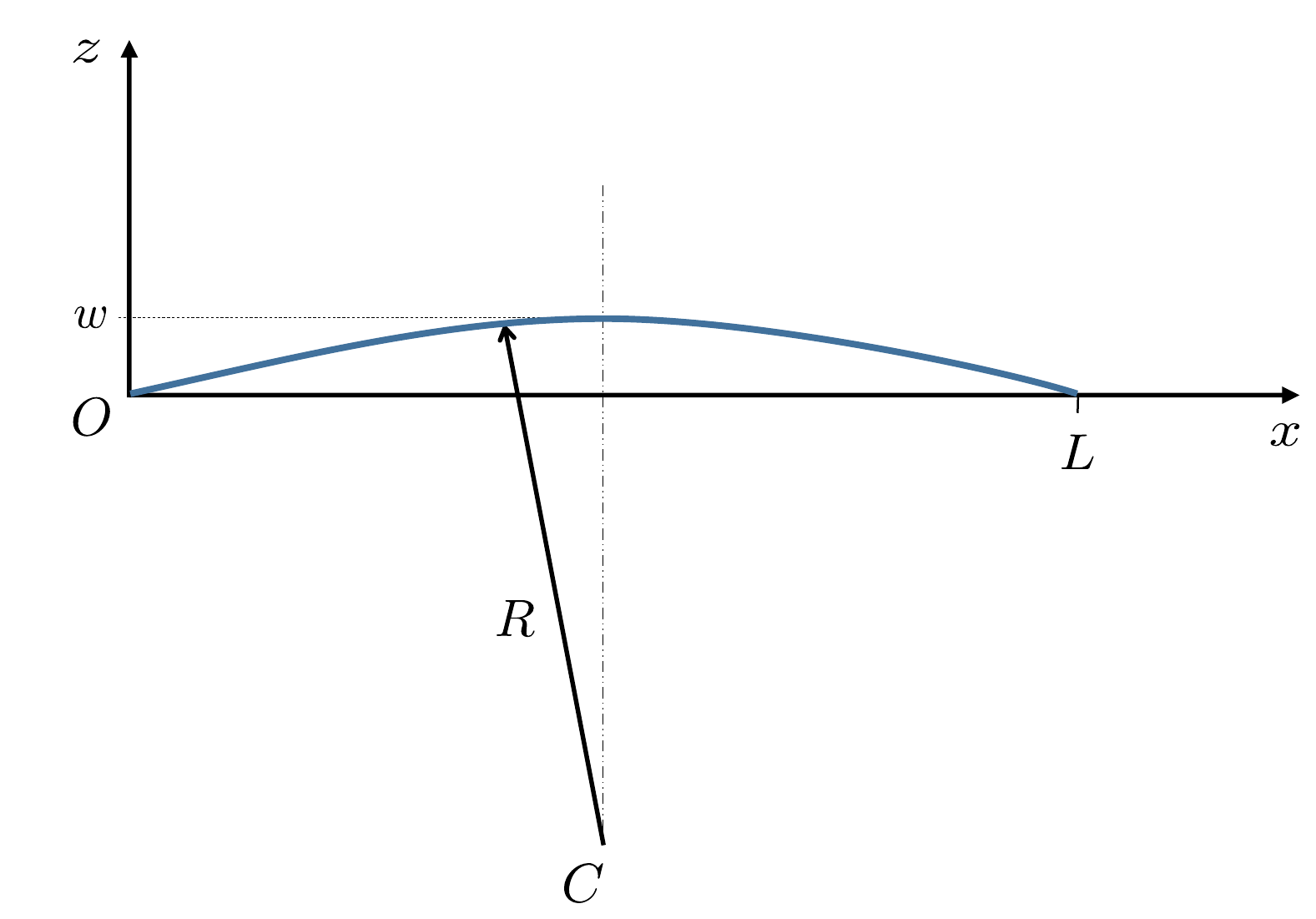}
		\caption{}
	\end{subfigure}
	\hspace{3mm}
	\begin{subfigure}{0.48\textwidth}
		\vspace{1cm}
		\centering\includegraphics[width=\linewidth]{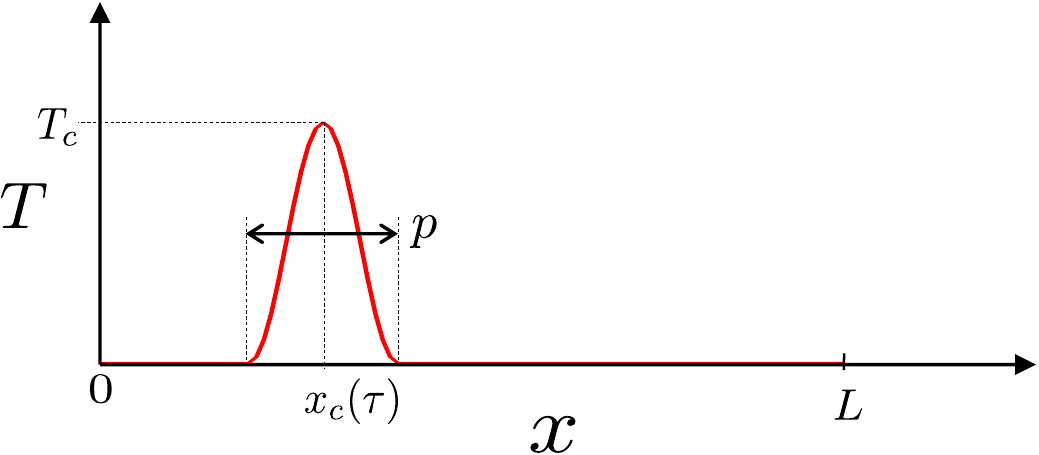}
		\vspace{0.5cm}\caption{}
	\end{subfigure}
	\caption{\label{fig:sketch}(\emph{a})\small An undeformed beam with length $L$
		and Radius of curvature $R$ is shown. The beam would be straight
		in the undeformed configuration if the curvature parameter $w=0$.
		(\emph{b}) A temperature pulse with spatial distribution given in eq.~\eqref{eq:temp_dist}}
\end{figure}
A pulse-shaped temperature field is applied to the beam, as shown in Figure~\ref{fig:sketch}.
It is modeled as using the following equation:
\begin{equation}
\label{eq:temp_dist}
T(x)=T_{c}\sin^{2}\left(\pi \frac{x-x_0}{p}\right)\left[H(x-x_0)-H\left((x-x_0) - p\right)\right],
\end{equation}
where $ x_0 = x_c - p/2 $, $ x_c $ is the pulse-center location, $ p $ is the width of the pulse, $ T_c $  is the pulse height, $ H $ is the Heaviside step function.
The center of the pulse $x_{c}$ moves slowly across the beam according
to the relation 
\begin{equation}
\label{eq:thermal_dynamics}
x_{c}=x_{0}+A\sin(\tau),
\end{equation}
where $x_{0}$ is the initial location of the center of the pulse
and $A$ is the amplitude of temperature-pulse-oscillation along the length of the beam.

\begin{table}[H]
	\begin{centering}
		\begin{tabular}{lr}
			\hline 
			Parameters (symbol) & Value {[}unit{]}\tabularnewline
			\hline 
			\hline 
			Length of beam ($L$) & 0.1 {[}m{]}\tabularnewline
			Thickness of beam ($h$) & 1 {[}mm{]}\tabularnewline
			Width of beam ($b$) & 10 {[}mm{]}\tabularnewline
			Curvature parameter for the curved- & \tabularnewline
			beam ($w$; cf. Figure \ref{fig:sketch}a; $ w=0 $ for straight beam) & 5 {[}mm{]} \tabularnewline
			\hline 
			Young's Modulus ($E$) & 70 {[}GPa{]}\tabularnewline
			Material damping modulus ($\kappa$) & 0.1 (GPa s)\tabularnewline
			Density ($\rho$) & 2700 {[}kg/m$^{3}${]}\tabularnewline
			Coefficient of linear expansion ($\alpha_{T}$) & 23.1$\times10^{-6}$ {[}K$^{-1}${]}\tabularnewline
			\hline 
		\end{tabular}
		\par\end{centering}
	\caption{\label{tab:props}Values for geometrical parameters and physical properties used in the beam models}
\end{table}
\begin{figure}[H]
	\centering
	\flushleft
	\begin{subfigure}{0.49\textwidth}
		\centering\includegraphics[width=\linewidth]{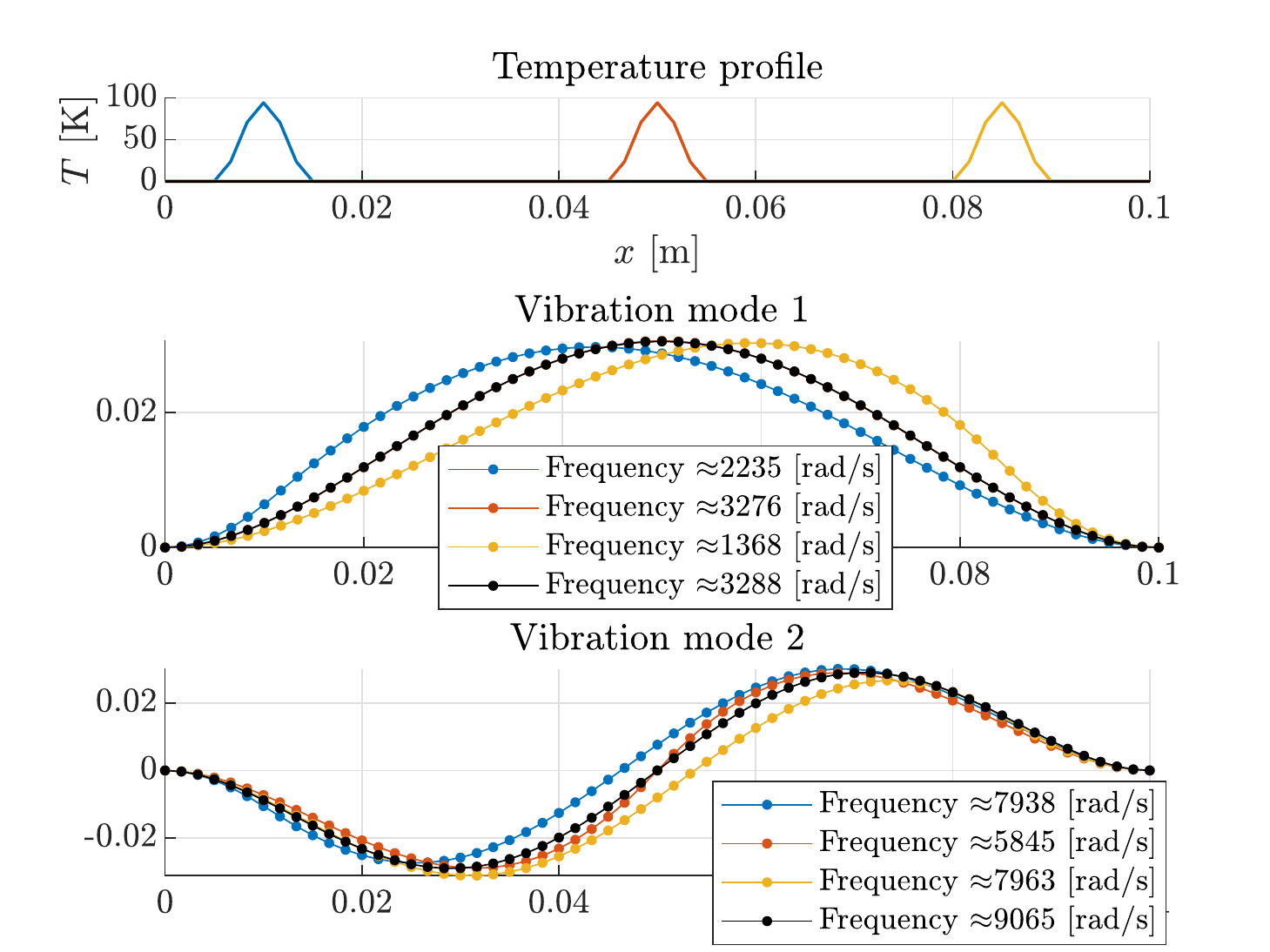}
		\caption{\small Initially straight beam}
	\end{subfigure}
	\begin{subfigure}{0.49\textwidth}
		\centering\includegraphics[width=\linewidth]{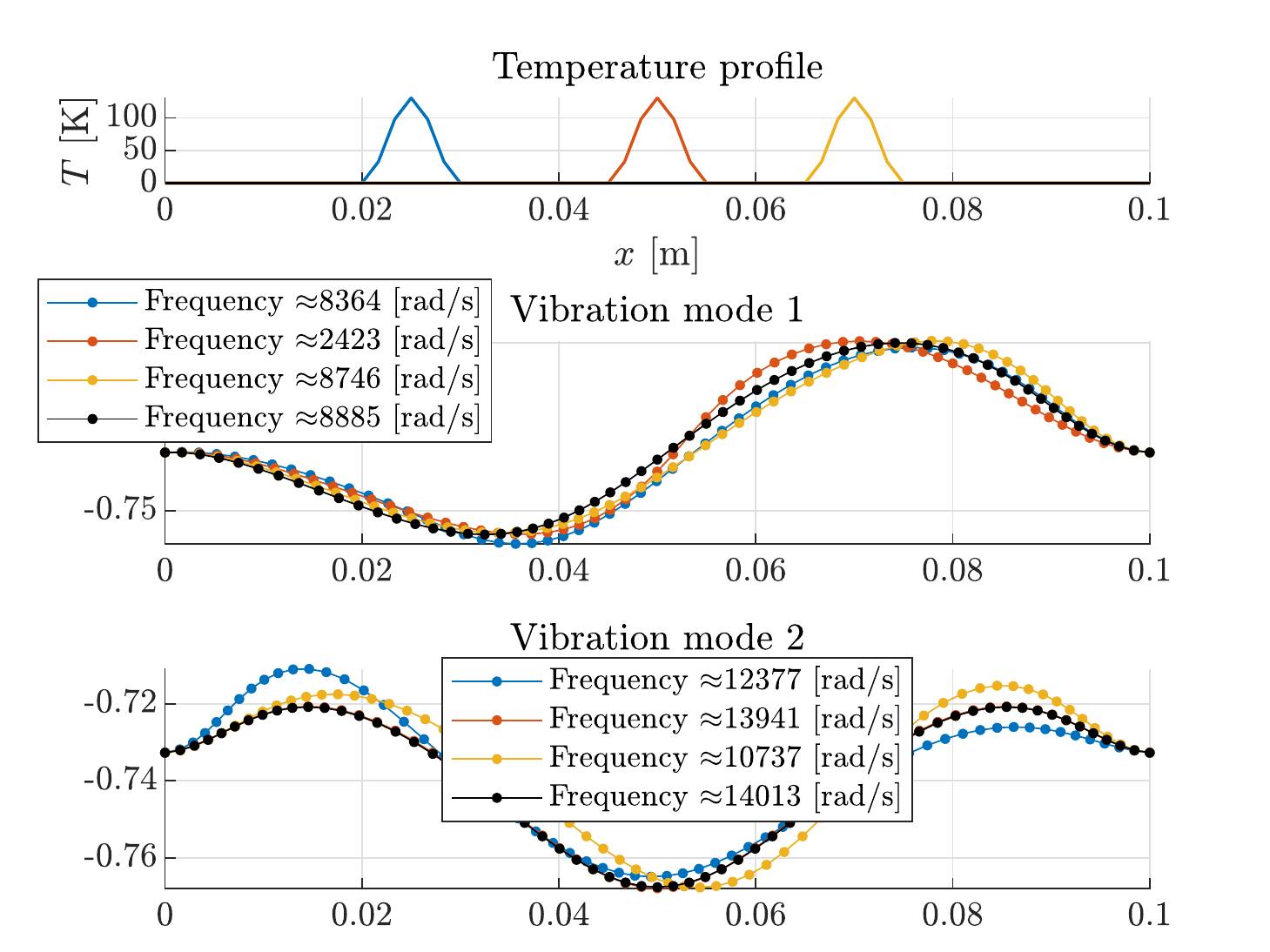}
		\caption{\small Intially curved beam}
	\end{subfigure}
	\caption{\label{fig:modes} \small The variation in the first two VMs of the doubly-clamped (a) straight and (b) curved beams (cf. Table~\ref{tab:props} for properties) with respect to the location of the temperature profile modeled according to eq.~\eqref{eq:temp_dist}, along with unheated or \emph{cold} configuration shown in black color. The mode shapes and corresponding frequencies change significantly across the three instances of the temperature profiles depicted. }
\end{figure}
We consider two geometries of the doubly-clamped beam model: one which is initially straight and the other with a small initial curvature. The FE model of the beam contains a temperature-dependent linear stiffness matrix. As a result, the VMs are also temperature-dependent. The variation in the linear VMs and natural frequencies of the beam for both models as the center of the temperature profile ($ x_c $) changes location are depicted in Figure~\ref{fig:modes}. 

Along with the imposed thermal dynamics on the beam, we also excite the structure mechanically with a load oscillating at a much faster rate than that of the thermal distribution, the details which are described in the following examples. 
Across all examples, we perform numerical time integration starting with the static equilibrium at the initial thermal configuration.

\textbf{Model reduction strategy:} During model reduction using the proposed MMS-based technique, a basis comprising of temperature-dependent VMs (and modal derivatives in case of nonlinear problems) is interpolated as the center $ x_c $ slowly changes according to eq.~\eqref{eq:thermal_dynamics}. In order to perform this interpolation, we use a database of $ n_d= 19$ bases, each computed around the equilibrium when the thermal pulse center is at location \begin{equation}\label{eq:temp_configs}
x_c^{(j)} = \frac{jL}{20}\,,\quad  j=1,\dots,19.
\end{equation} We ensure that all the bases in this database are consistent using Algorithm~\ref{alg:congruent_basis}. As the temperature profile moves along the length of the beam, we obtain a basis relevant for the instantaneous temperature configuration by performing  linear interpolation between the bases in the database. In general, such an interpolation is not guaranteed to preserve orthogonality of the bases and can even lead to ill-conditioning. This can be avoided by interpolating the bases over a Stiefel manifold, as proposed in ref.~\cite{Amsallem_2009}. However, in the authors' experience, this approach is not suitable for performing interpolation online, as it can be computationally intensive and a linear interpolation is found to work just as well in structural dynamics applications.


\textbf{Comparison with a stacking-based approach}: A simple and robust approach for reduction would constitute stacking the modes relevant to different temperature configurations in a single basis and performing orthogonalization, e.g., using the Gram-Schmidt method or singular value decomposition (SVD) (cf. Golub \& van Loan~\cite{Golub2013}) . Specifically, let $ \mathbf{V}^{(j)} \in\mathbb{R}^{n\times m} $ be an orthonormal basis for the $ j^{\mathrm{th}} $ temperature configuration in the database of $ n_d $ configurations. We stack these bases in a matrix \begin{equation}
	\label{eq:Vdef}
	\boldsymbol{\mathcal{V}} :=[\mathbf{V}^{(1)},\dots,\mathbf{V}^{(n_d)}]\in \mathbb{R}^{n\times n_dm}
\end{equation} 
to perform SVD as
	\begin{equation}
	\label{eq:svd}
		\boldsymbol{\mathcal{V}} = \mathbf{L} \mathbf{\Sigma} \mathbf{R}^{\top}, 
	\end{equation}	
where $ \mathbf{L}\in \mathbb{R}^{n\times n} $ is an orthonormal matrix containing the left singular vectors, $ \mathbf{\Sigma} \in \mathbb{R}^{n\times n_d m }$ is a diagonal matrix containing the corresponding singular values, and $ \mathbf{R}\in\mathbb{R}^{n_d m\times n_dm} $ is an orthonormal matrix containing the right singular vectors. The columns of $ \mathbf{L} $ corresponding to the non-zero singular values in $ \mathbf{\Sigma} $ form an orthonormal basis spanning the columns of $ \boldsymbol{\mathcal{V}} $. Following this procedure, the $ \textsc{matlab} $ command $ \texttt{orth}(\boldsymbol{\mathcal{V}}) $ economically computes and returns the required orthonormal basis that we have used in this work.\footnote{Arguably, such a basis is conservative and might contain spurious directions due to numerical tolerances. A workaround to avoid such spurious modes would be to consider the modes associated to the non-zero singular values greater than a chosen numerical threshold, which is considerably higher than the machine zero value.}  

\paragraph{Remark.} The usual approach of plotting the singular value decay to cut-off the mode selection after appearance of a ``knee" in such a plot is particularly justifiable in the context of POD, where a dominant subspace or principal components are identified from simulation snapshots or data.  Such simulation snapshots carry trajectory information in contrast  to the reduced bases that already represent dominant direction at different parameter configurations in the matrix $ \boldsymbol{\mathcal{V}} $. In other words, one needs a basis that spans all the columns of $ \boldsymbol{\mathcal{V}} $ as opposed to obtaining "dominant directions" that approximately span the data contained in $ \boldsymbol{\mathcal{V}} $.

\begin{figure}[H]
	\centering
	\begin{subfigure}{0.32\textwidth}
		\centering\includegraphics[width=\linewidth]{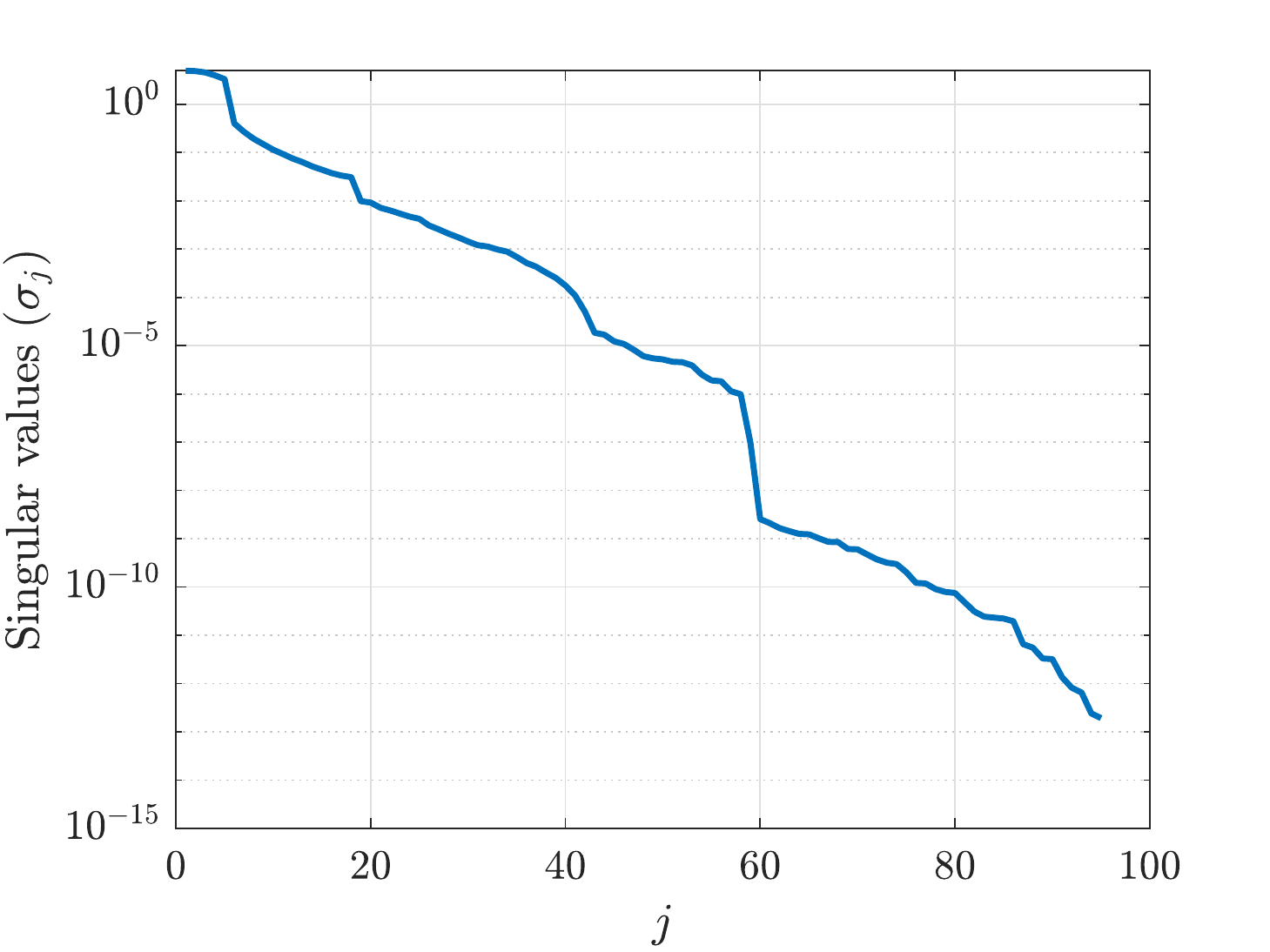}
		\caption{\small Straight beam (VMs)}
	\end{subfigure}
	\begin{subfigure}{0.32\textwidth}
		\centering\includegraphics[width=\linewidth]{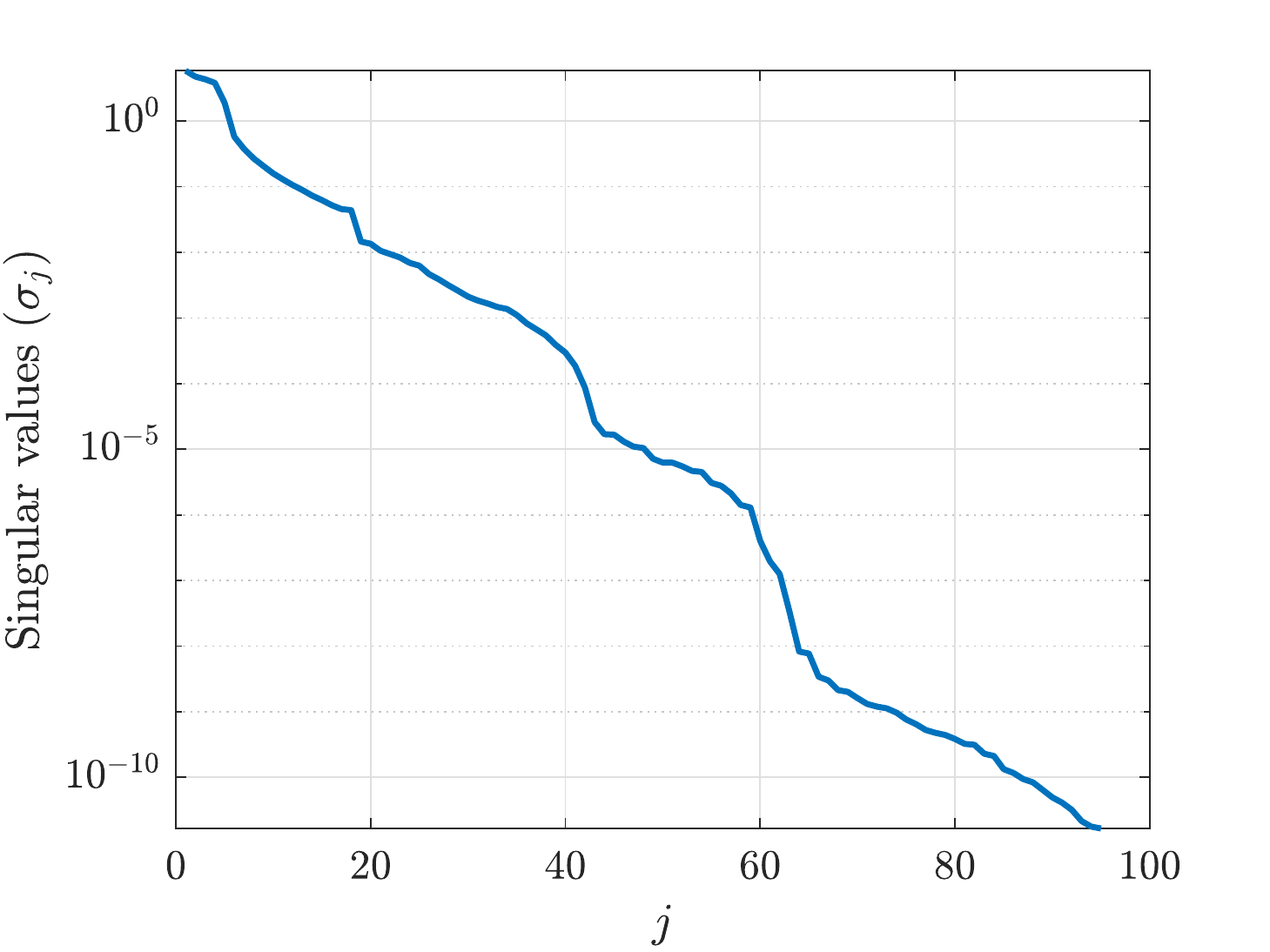}
		\caption{\small Curved beam (VMs)}
	\end{subfigure}
	\begin{subfigure}{0.32\textwidth}
		\centering\includegraphics[width=\linewidth]{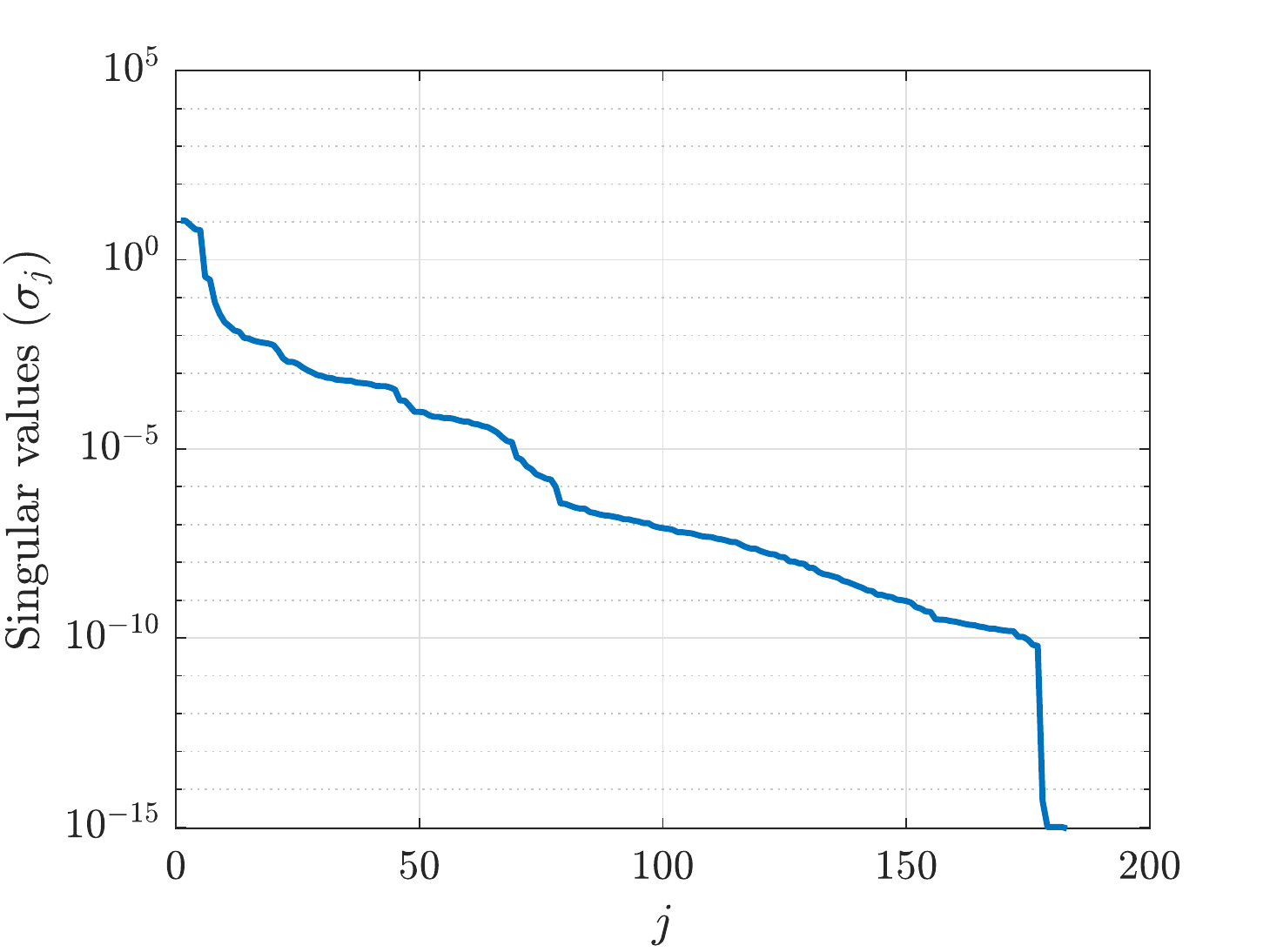}
		\caption{\small Curved beam (VMs \& modal derivatives )}
	\end{subfigure}
	\caption{\label{fig:svd} \small The decay of singular values of the matrix $ \boldsymbol{\mathcal{V}} $ (cf.eq.~\eqref{eq:Vdef}) that consists reduction bases at different temperature configurations. The bases are composed of the first five vibration modes (and the corresponding modal derivatives in case of (c)) for (a) straight beam, (b) \& (c) curved beam. The figure shows that temperature (parameter) change may lead to a significant change of modes that cannot be captured effectively using a low dimensional basis.}
\end{figure}

The orthogonalized basis is expected to capture behavior across a range of temperature configurations in a more robust manner in comparison with other approaches mentioned in the Introduction. However, the size of such a basis can very quickly become unmanageable as more bases at different temperature configurations are added to the database, as shown by Figure~\ref{fig:svd}. Furthermore, note that our simple examples feature a single parameter for basis description, i.e., the location $ x_c $ of the center of a temperature pulse. This simple stacking-based approach may lead to a further increase in basis size as the parameter-dimensionality increases. To avoid such high-dimensionality for a reduced-basis, we adopt the following methods to limit the size of the reduction basis used for comparison purposes, when necessary:
\begin{enumerate}	
	\item \textbf{Modal-POD}: Here, we use $ m $ modes associated to the highest singular values obtained from the SVD~\eqref{eq:svd} of the matrix $ \boldsymbol{\mathcal{V}} $ given in eq.~\eqref{eq:Vdef}. This results in the best constant basis of size $ m $ that can be obtained from the temperature configurations in the database. Note that the size $ m $ is chosen to be the same as that of the reduction basis used in the MMS-based reduction approach to facilitate a fair comparison in terms of the number of unknowns.
	\item \textbf{Modal}: One might wish to avoid the Modal-POD type reduction in view of the above-mentioned remark. In that case, one could select a subset of temperature configurations at random from the $ n_d $ configurations available in the database. Thus, lesser data is used to construct the matrix $ \boldsymbol{\mathcal{V}} $ and a smaller basis size is obtained after orthogonalization.
\end{enumerate}

\subsection{Straight beam}
First we consider a linear model of a straight beam, whereby the nonlinear terms arising from the von-K\'{a}rm\'{a}n assumption are neglected. Note, however, that the model still contains a temperature-dependent linear stiffness matrix along with a thermal load vector.  In addition, the beam is mechanically excited as follows:
\begin{equation}
\label{eq:mech_load}
	\mathbf{p}(t,\epsilon)=\mathbf{l}\sin\omega_s t,
\end{equation}
where $\omega_s$ is the loading frequency, chosen to be the average
of the first and second natural frequency (calculated for temperature configuration $ x_c = L/2 $), and $ \mathbf{l} $ is load amplitude vector, representing a spatially uniform load in the transverse direction with a density of $ 10^4 $N/m. The temperature profile  traverses the beam span according to the relation~\eqref{eq:thermal_dynamics} with $ x_0 = 0.5L, A=0.3L $.  

Using the stacking approach described above that involves stacking all the modes in the database together, we obtain a basis of size (rank) 93, which is quite large for full system with 177 DOFs. It turns out that this simple problem can be reduced using a constant basis of five modes according to the Modal-POD described above. We compare the solution using the MMS-based reduction approach proposed in this work using first five VMs which are adapted to the instantaneous temperature configuration. We test our approach for $ \epsilon =0.01,0.001$.
\begin{figure}[H]
	\centering
	\begin{subfigure}{\textwidth}
		\centering\includegraphics[width=\linewidth]{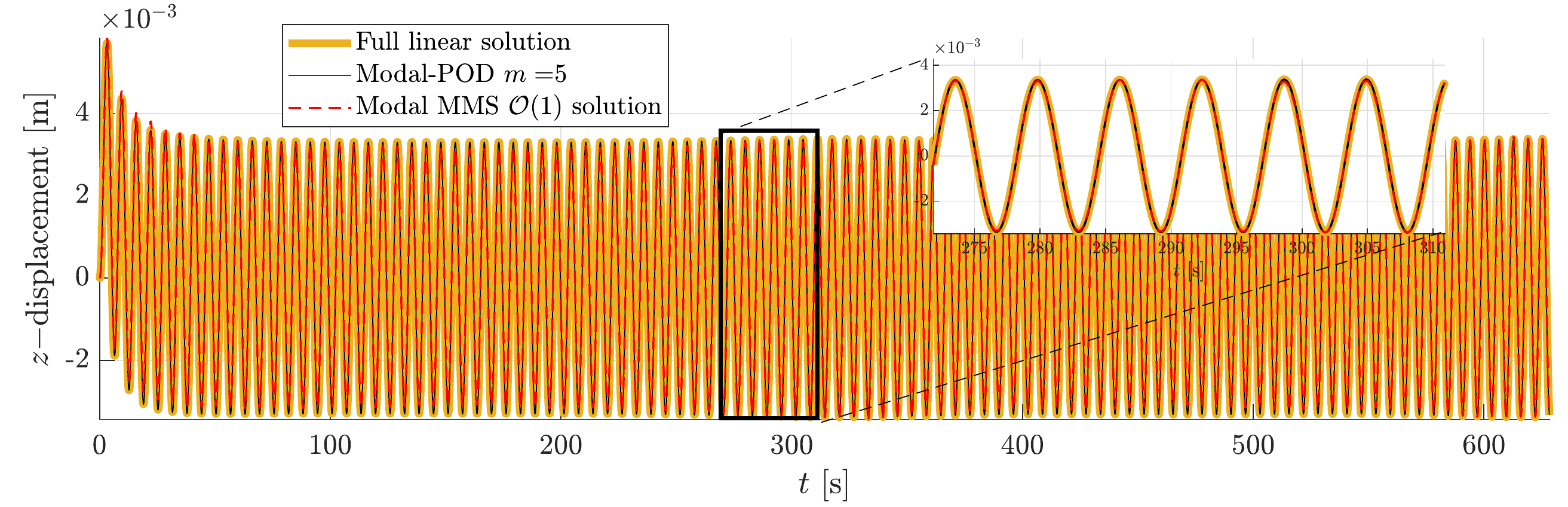}
		\caption{The reduced solution using a constant basis of 5 VMs (Modal-POD, solid black line) is just as effective in approximating transverse displacements as the leading order reduced solution (broken-red line) using the MMS.}
	\end{subfigure}
	
	\vspace{5mm}
	
	\begin{subfigure}{\textwidth}
		\centering\includegraphics[width=\linewidth]{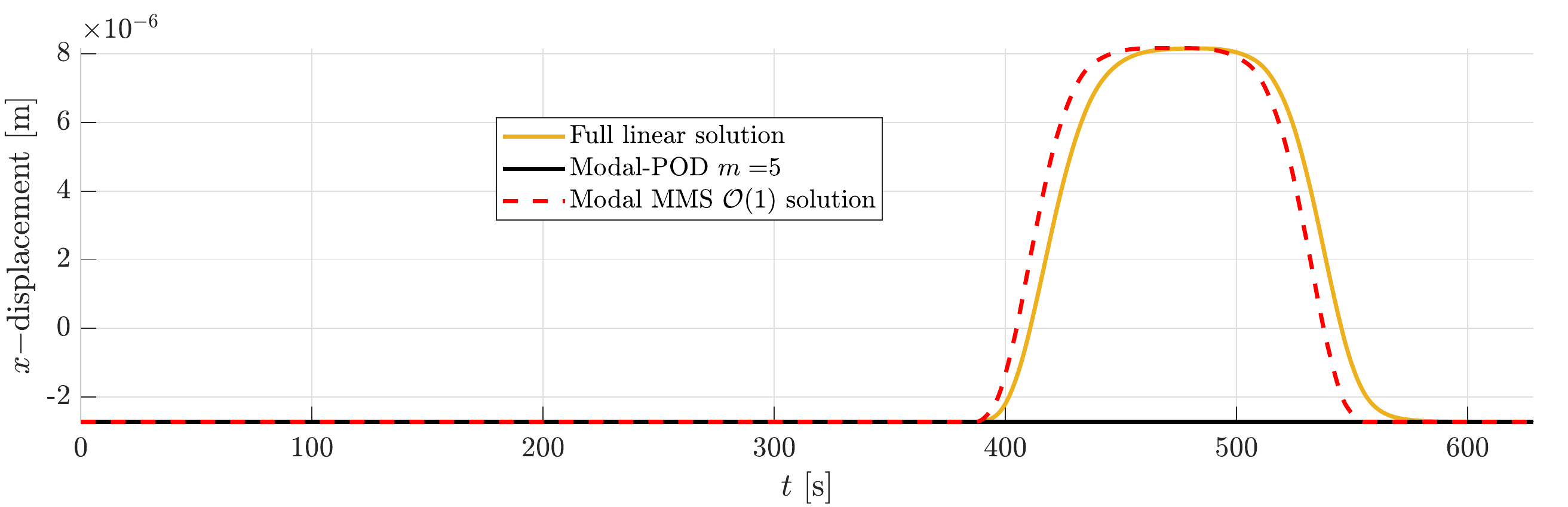}
		\caption{The axial displacements are expectedly missed by the reduced solution using a constant basis of Modal-POD ($ m=5 $) shown by the solid black line. The MMS-based reduced solution (broken-red line) captures the qualitative behavior of the full solution. }
	\end{subfigure}
	\caption{\small \label{fig:strt_lin} The (a) transverse and (b) axial displacements for the node situated at $ x=L/4 $ in response to combined thermo-mechanical loading \eqref{eq:mech_load}, \eqref{eq:thermal_dynamics} with $ \epsilon=0.01 $. The simulation is performed for 100 cycles of mechanical loading when the temperature profile is traversing along the length of the beam. Note that the time on the horizontal axis is non-dimensionalized and rescaled using the time-period of loading such that each loading cycle corresponds to $ 2\pi $ units on the time axis.}
\end{figure}
The results show that the former approach using a constant basis is just as effective in capturing transverse displacements of the beam as the multiple-scales approach using an adaptive basis (cf. Figure~\ref{fig:strt_lin}a). Note, however, that the axial displacements are totally missed by this approach, as shown in Figure~\ref{fig:strt_lin}b. This is not unexpected since the first five VMs contain only low-frequency bending fields. Such a basis would thus be ineffective in capturing the membrane effects. On the other hand, the MMS-based approach systematically takes care of this issue while still using a similar basis which contains bending modes only. This is achieved by making use of the scale separation when axial components are systematically captured using a temperature-dependent equilibrium according to eq.~\eqref{eq:leading_order_linear}. For $ \epsilon=0.01 $, while the MMS-based approach qualitatively captures the axial response, it is not as accurate as in capturing the transverse displacements (cf. Figure~\ref{fig:strt_lin}). However, the full system response, including the axial displacements is expected to be captured accurately for sufficiently small $ \epsilon $. Indeed, this is the case for a smaller value of $ \epsilon=0.001 $,  as shown in Figure~\ref{fig:strt_lin_3}.
\begin{figure}[H]
		\centering\includegraphics[width=\linewidth]{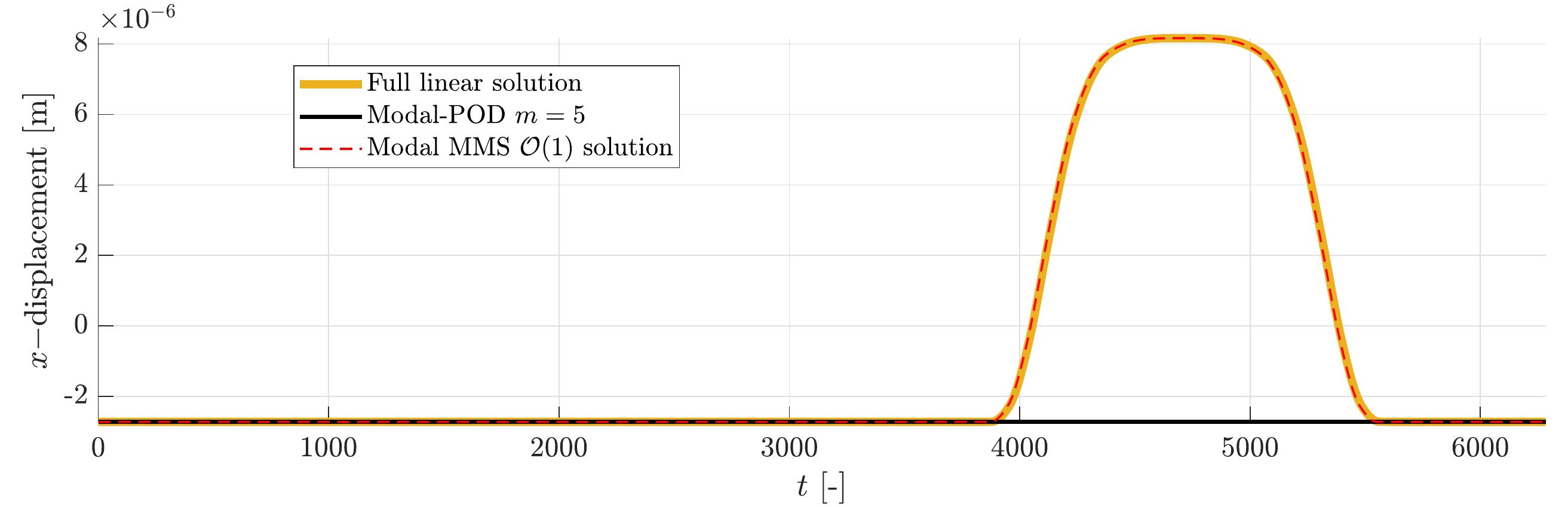}
		
		\caption{ \label{fig:strt_lin_3}Same as Figure~\ref{fig:strt_lin}, except that the thermal loading is applied at a slower rate($ \epsilon=0.001 $). In contrast to Figure~\ref{fig:strt_lin}b, the axial displacements are very accurately captured using the MMS-based reduction for this smaller epsilon. The simulation is performed for 1000 cycles of mechanical loading.  }
\end{figure}

Instead of using the proposed MMS-based reduction, the axial contributions can also be captured by appending a set of carefully selected axial modes in the original constant basis (Modal-POD). Arguably, this would be a better way to reduce the problem because even though we increase the basis size, it completely avoids the basis interpolation required in the MMS-based approach. While selecting such axial modes is relatively easy in this illustrative straight beam example, this could already be challenging if we add curvature to the beam, thereby coupling the bending and membrane DOFs. The MMS-based approach is still as effective and efficient in such situations, as we show in the next example.

\subsection{Curved beam}

\begin{figure}[H]
	\centering
	\begin{subfigure}{0.49\linewidth}
		\centering\includegraphics[width=\linewidth]{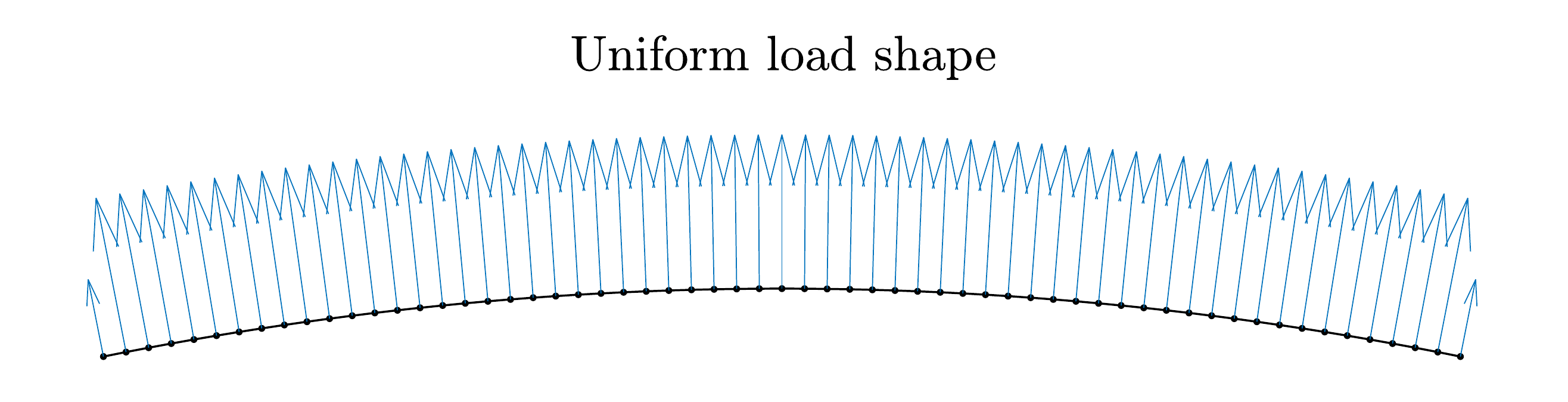}
		\caption{\small Shape of field $ \mathbf{l}_0 $ in eq.~\eqref{eq:pert_forcing}}
	\end{subfigure}
	\begin{subfigure}{0.49\linewidth}
		\centering\includegraphics[width=\linewidth]{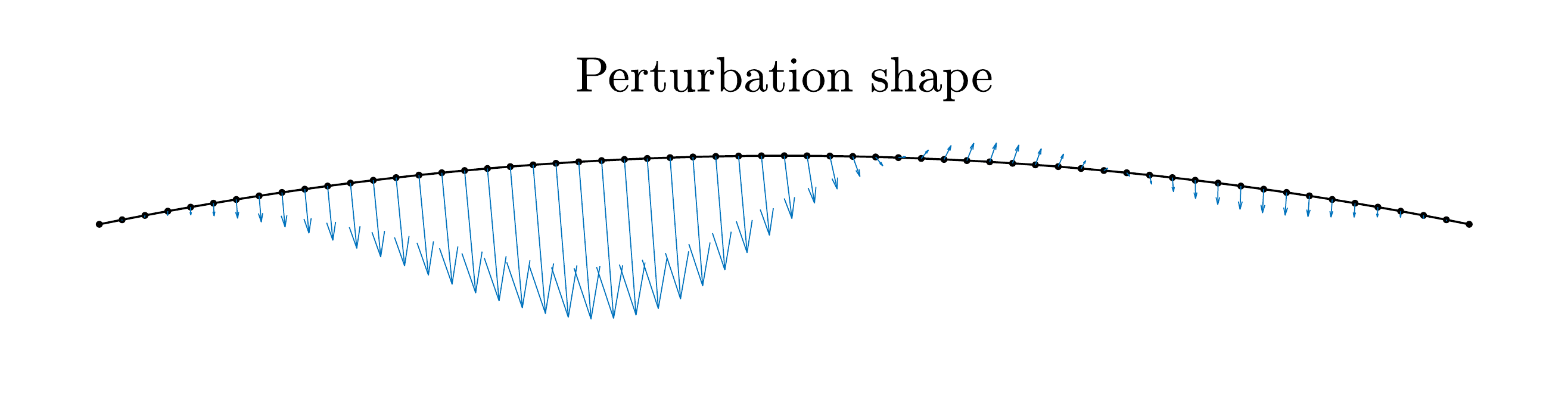}
		\caption{\small Shape of field $ \mathbf{l}_1 $ in eq.~\eqref{eq:pert_forcing}}
	\end{subfigure}
	\begin{subfigure}{\textwidth}
		\centering\includegraphics[width=\linewidth]{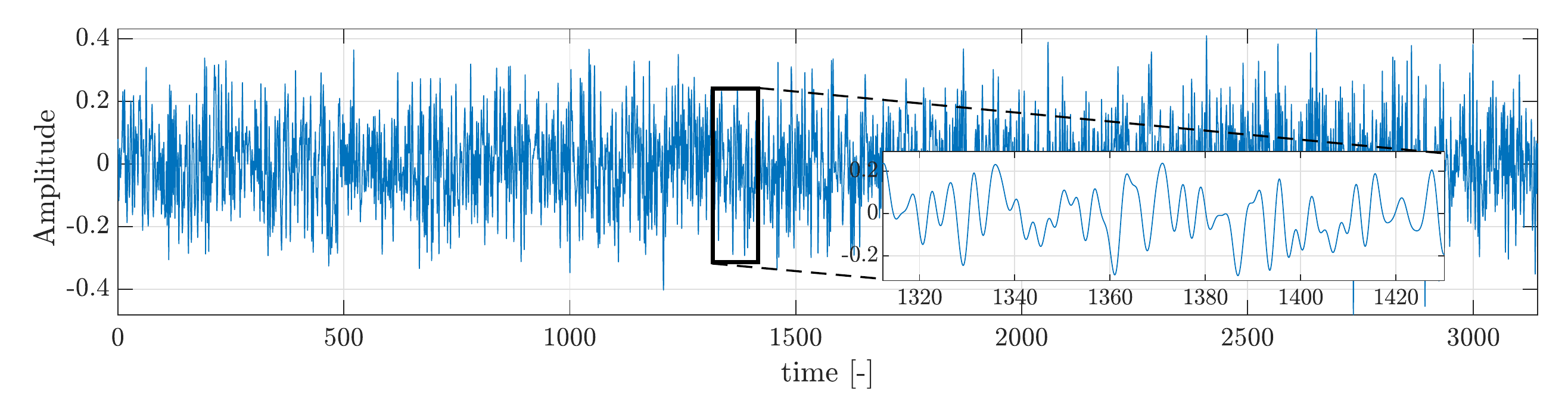}
		\caption{\small Amplitude $ a $ in eq.~\eqref{eq:pert_forcing} for 500 cycles of leading order loading at frequency $ \omega_c $ rad/s. Note that the time on the horizontal axis is rescaled using the time-period $ 2\pi/\omega_c $ such that each loading cycle corresponds to $ 2\pi $ units on the time axis.}
	\end{subfigure}	
	\caption{ \label{fig:randper} \small Description of the loading on the curved beam (cf. Figure~\ref{fig:sketch}, Table~\ref{tab:props}) according to eq~\eqref{eq:pert_forcing}.  }
\end{figure}

Next, we consider a beam with the same properties as the previous example but curvature added (cf. Table~\ref{tab:props}). We consider the following mechanical excitation applied to the beam
\begin{equation}
\mathbf{g}(t) = \mathbf{l}_0\sin\omega_c t + \tilde{\mathbf{l}}_1 a(t)\,,
\end{equation}
where $\omega_c$ is a typical loading frequency, once again taken as the average
of the first and second natural frequency of the system (calculated for the temperature configuration $ x_c = L/2 $); $ \mathbf{l}_0 $ is the spatial load vector at the leading order whose shape represents a uniform pressure at the magnitude of $ 10^3 $ N/m (cf.~Figure~\ref{fig:randper}a for its shape). In addition to this uniform load, we have a random $ \mathcal{O}(\epsilon) $ perturbation that is modeled as follows. The perturbation shape $ \tilde{\mathbf{l}}_1 $ (see~Figure~\ref{fig:randper}b) is constructed by a linear combination of the first five VMs of the beam (computed for $ x_c=L/2 $) such that the coefficients for this combination are pseudo-random values from the standard uniform distribution in the interval $ (0,1) $. The time variation of this perturbation is given by the scalar function $ a(t) $ which is modeled by choosing pseudo-random values at each time instant from the uniform distribution in the interval (0,1) and thereafter filtering out the frequency contents higher than the third natural frequency of the beam. The resulting amplitude has time history as shown in ~Figure~\ref{fig:randper}c.

Finally, we assume that the random perturbation $ \tilde{\mathbf{l}}_1$ models noise and has small amplitude in comparison with the leading-order forcing vector $ \mathbf{l}_0$. As discussed in Section~\ref{sec:setup}, we scale this given component of the forcing as $ \tilde{\mathbf{l}}_1 = \epsilon\mathbf{l}_1  $, where $ \mathbf{l}_1:= \tilde{\mathbf{l}}_1/\epsilon $. Our choice of loading amplitudes ensures that $ \mathbf{l}_0 $ and $ \mathbf{l}_1 $ are of similar magnitudes after this scaling, and in particular $ \|\mathbf{l}_1\|_2 = \|\mathbf{l}_0\|_2 $. We obtain
\begin{equation}
\label{eq:pert_forcing}
\mathbf{g}(t) = \mathbf{p}(t,\epsilon)=\mathbf{l}_0\sin\omega_c t + \epsilon \mathbf{l}_1 a(t).
\end{equation}

\subsubsection{Linear model}
\begin{figure}[H]
	\centering
	\begin{subfigure}{\textwidth}
		\centering\includegraphics[width=\linewidth]{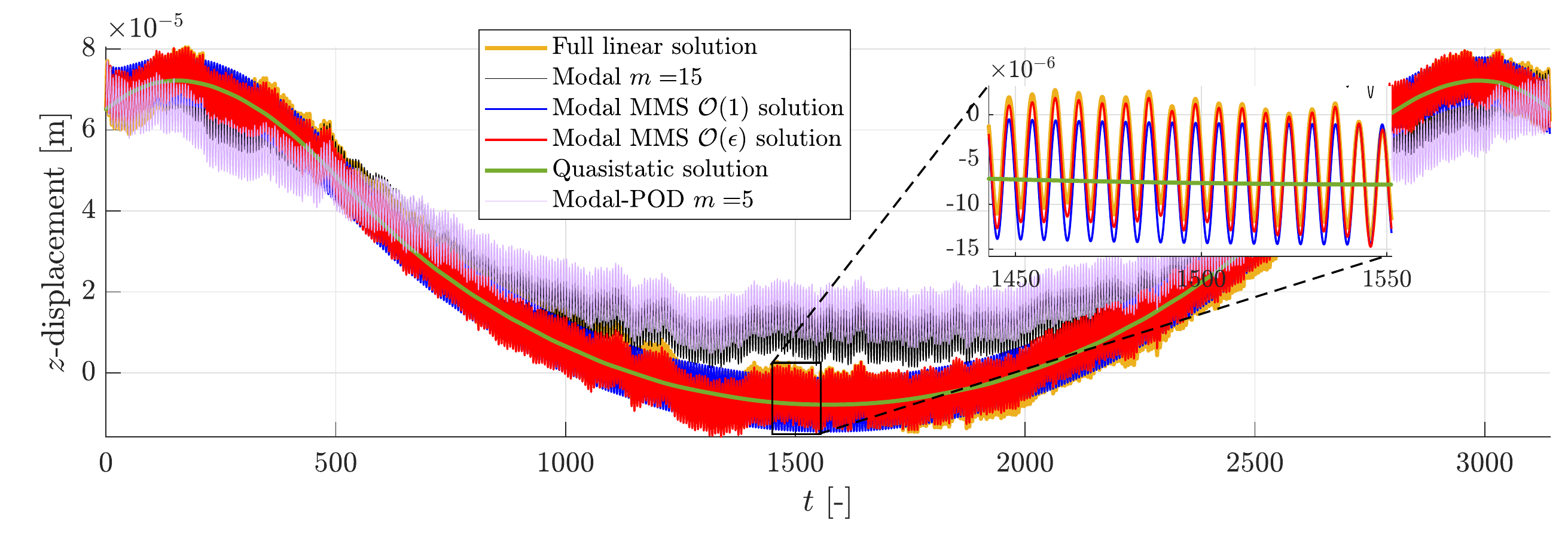}
		\caption{ }
	\end{subfigure}
	\begin{subfigure}{\textwidth}
		\centering\includegraphics[width=\linewidth]{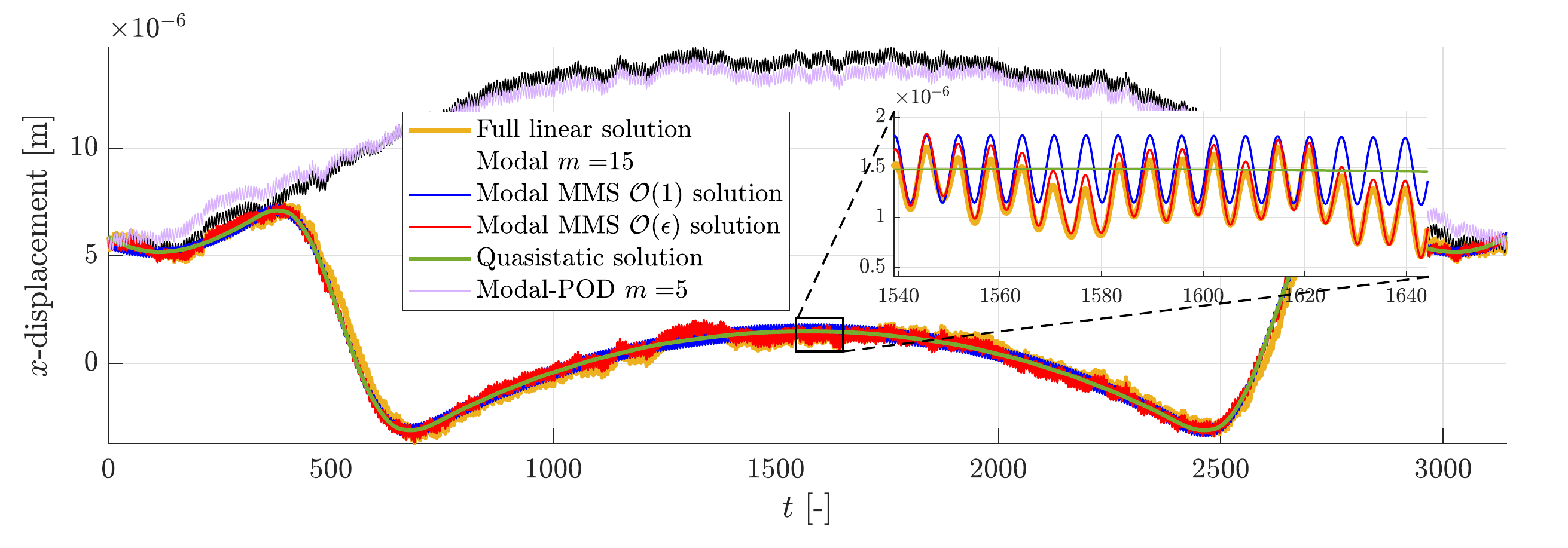}
		\caption{}
	\end{subfigure}	
	\caption{ \label{fig:curved_lin_rand} \small The (a) $ z$-direction and (b) $ x$-direction displacements for the node situated at $ x=L/4 $ in response to combined thermo-mechanical loading \eqref{eq:mech_load} (cf.~Figure~\ref{fig:randper}), \eqref{eq:thermal_dynamics} with $ x_0 = 0.1L, A=0.3L, \epsilon=10^{-3} $. The Full system solution is depicted in the solid yellow line; the solid black line (Modal $ m=15 $) shows the reduced solution obtained from Galerkin projection onto a constant basis containing 15 modes; the solid blue (Modal MMS $ \mathcal{O}(1) $) and red (Modal MMS $ \mathcal{O}(\epsilon) $) lines show the reduced solution using the MMS-based reduction upto $ \mathcal{O}(1)  $ and $ \mathcal{O}(\epsilon) $ accuracy respectively, using a basis of $ m=5 $ modes adapting to the instantaneous temperature configuration; solid pink line (Modal-POD $m=5$) shows the reduced solution obtained from Galerkin projection onto a constant basis containing 5 modes with the highest singular values in from eq.~\eqref{eq:svd}. }
\end{figure}

We first consider the linear example of the curved beam with the mechanical forcing~\eqref{eq:pert_forcing} along with a time-varying temperature distribution according to relation~\eqref{eq:thermal_dynamics}, as before.  Since the spectral content of the applied mechanical forcing is contained within the first five modes of the systems, these modes constitute an ideal choice for reduction. Thus, similar to the previous straight-beam example, we use the MMS-based reduction with a 5-mode basis that adapts instantaneously  to the temperature configuration of the beam; see Figure~\ref{fig:curved_lin_rand} for results when $ \epsilon=10^{-3} $. The presence of the $ \mathcal{O}(\epsilon) $ (pseudo-random) perturbation to the leading-order forcing, however, leads to more interesting results in this case. Note that the leading order ROM is uninfluenced by the $ \mathcal{O}(\epsilon) $ component of the forcing and misses the corresponding fluctuating components. Nonetheless, these components are successfully captured by the $ \mathcal{O}(\epsilon) $-ROM, as shown in Figure~\ref{fig:curved_lin_rand}.

On the other hand, the stacking-based approach described earlier results in a basis of size 95, which is again not feasible for reducing a 177 DOF system. To alleviate this:
\begin{enumerate}
	\item Modal ($ m=15 $): we apply the stacking approach to a subset of the sampled temperature configurations. The first five modes are available around the temperature configurations~$ x_c^{(j)} $ described in eq.~\eqref{eq:temp_configs}. We select 3 of these temperature configurations at random ($ j = 2,12,18 $ in this instance) and stack the corresponding modes in a matrix and perform orthogonalization to obtain a reduction basis of size 15.
	\item Modal-POD ($ m=5 $): we obtain the $ m=5 $ modes associated to the highest singular values of the matrix $ \boldsymbol{\mathcal{V}} $ which has $ n_d = 19 $ temperature configurations (cf.~eqs.~\eqref{eq:temp_configs},\eqref{eq:Vdef}).
\end{enumerate}

A conventional reduction is then performed via Galerkin projection using these bases. The corresponding reduced solutions show a significantly worse accuracy (see~Figure~\ref{fig:curved_lin_rand}) when compared to the MMS-based reduction at leading order, as well as at $ \mathcal{O}(\epsilon) $.

\begin{figure}[H]
	\centering
	\begin{subfigure}{\textwidth}
		\centering\includegraphics[width=\linewidth]{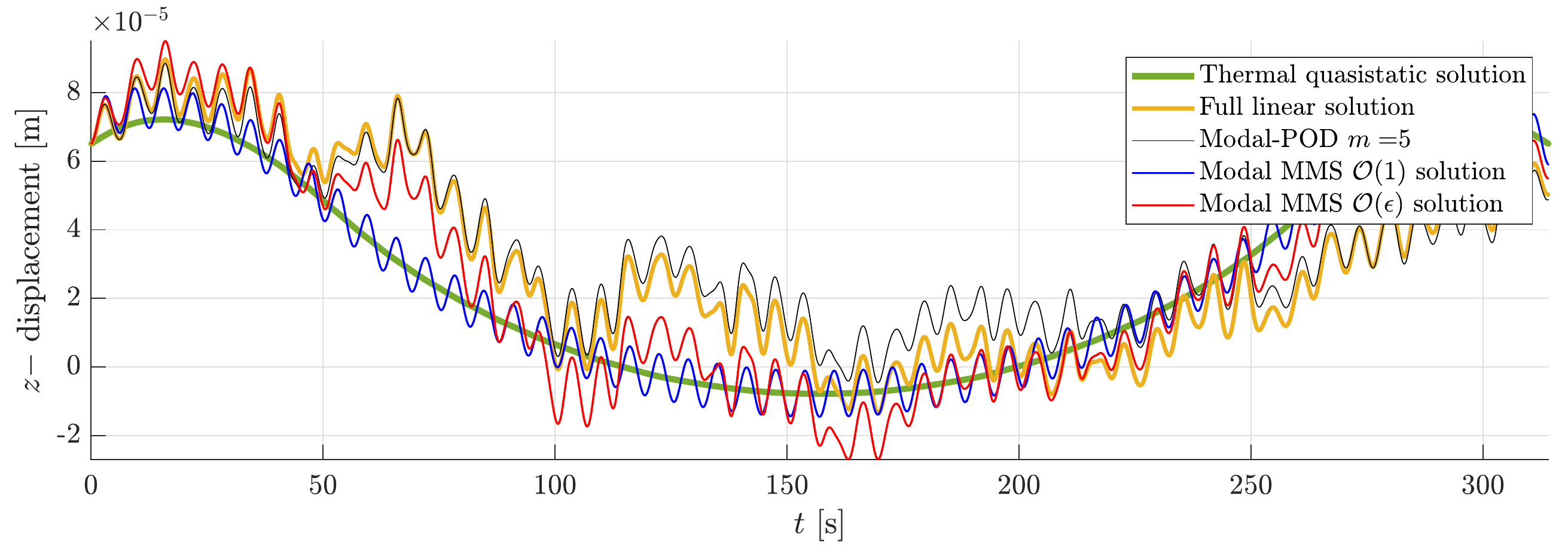}
		\caption{ }
	\end{subfigure}
	\begin{subfigure}{\textwidth}
		\centering\includegraphics[width=\linewidth]{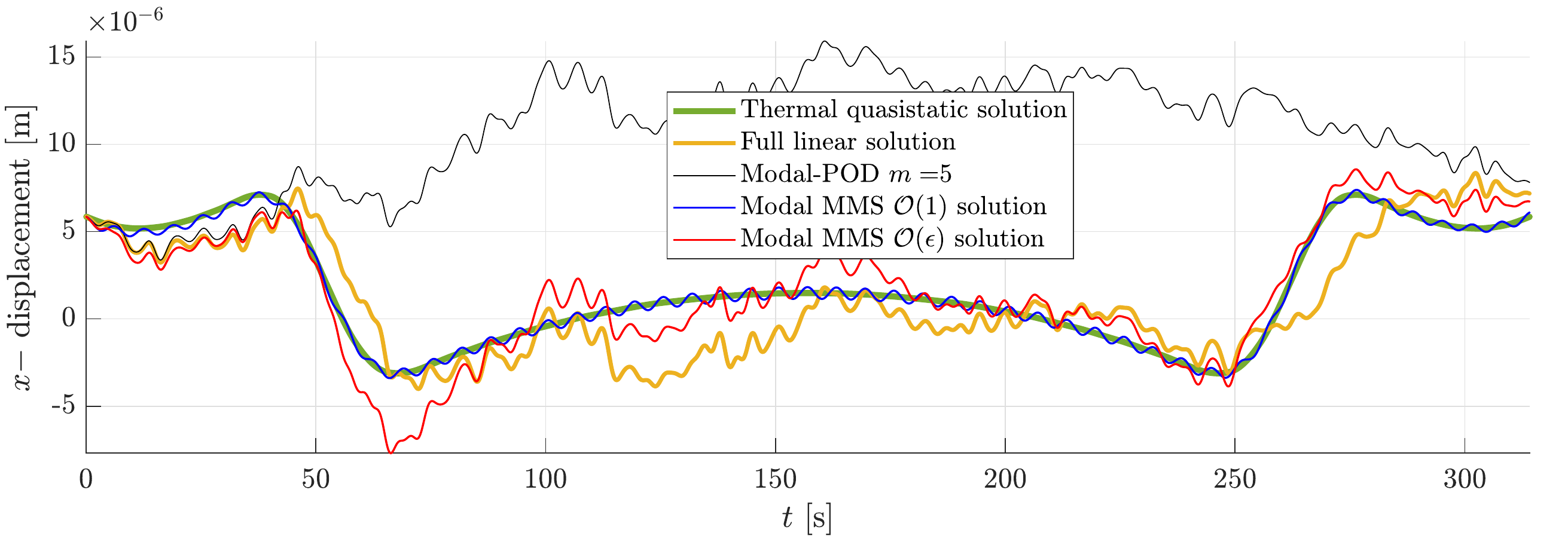}
		\caption{}
	\end{subfigure}	
	\caption{ \label{fig:curved_lin_rand_eps_0.01} \small The (a) $ z$-direction and (b) $ x$-direction displacements for the node situated at $ x=L/4 $ in response to combined thermo-mechanical loading \eqref{eq:mech_load} (cf.~Figure~\ref{fig:randper}), \eqref{eq:thermal_dynamics} with $ x_0 = 0.1L, A=0.3L, \epsilon=10^{-2} $. The Full system solution is depicted in the solid yellow line;  solid pink line (Modal-POD $m=5$) shows the reduced solution obtained from Galerkin projection onto a constant basis containing 5 modes with the highest singular values in from eq.~\eqref{eq:svd}; the solid blue (Modal MMS $ \mathcal{O}(1) $) and red (Modal MMS $ \mathcal{O}(\epsilon) $) lines show the reduced solution using the MMS-based reduction at $ \mathcal{O}(1)  $ and $ \mathcal{O}(\epsilon) $ using a basis of $ m=5 $ modes adapting to the instantaneous temperature configuration. }
\end{figure}

Upon increasing $ \epsilon $, which quantifies the speed of thermal loading, we observe that the multiple scales assumption starts to deviate from the true behavior of the system, as expected. The same is demonstrated in the introduction with the simple examples in Figure~\ref{fig:2dof_ROM_Tdyn}. In particular, for $ \epsilon=0.01 $, we see in Figure~\ref{fig:curved_lin_rand_eps_0.01}  that the MMS-based reduction approach is unable to capture the Full system response, even on this linear example. The same example also demonstrates the failure of the conventional Galerkin projection-based techniques to reduce the Full system using a constant basis.
\subsubsection{Nonlinear model}
Finally, we consider the above-described curved beam model, but this time with the presence of geometric nonlinearities. As discussed in Section~\ref{sec:MMS_nl}, it is well-known that a basis consisting solely of a few linear VMs is insufficient to capture the nonlinear response of a system, and the enrichment of the reduction basis with modal derivatives is an effective way to capture geometrically nonlinear response. Accordingly, we augment the database of 5-mode bases used in the linear case with the corresponding static modal derivatives. Since $ k=5 $ VMs result in $ k(k+1)/2 = 15 $ modal derivatives, we obtain a basis of size $ m=20 $ at each temperature configuration in the database. 

\begin{figure}[H]
	\centering
	\begin{subfigure}{\textwidth}
		\centering\includegraphics[width=\linewidth]{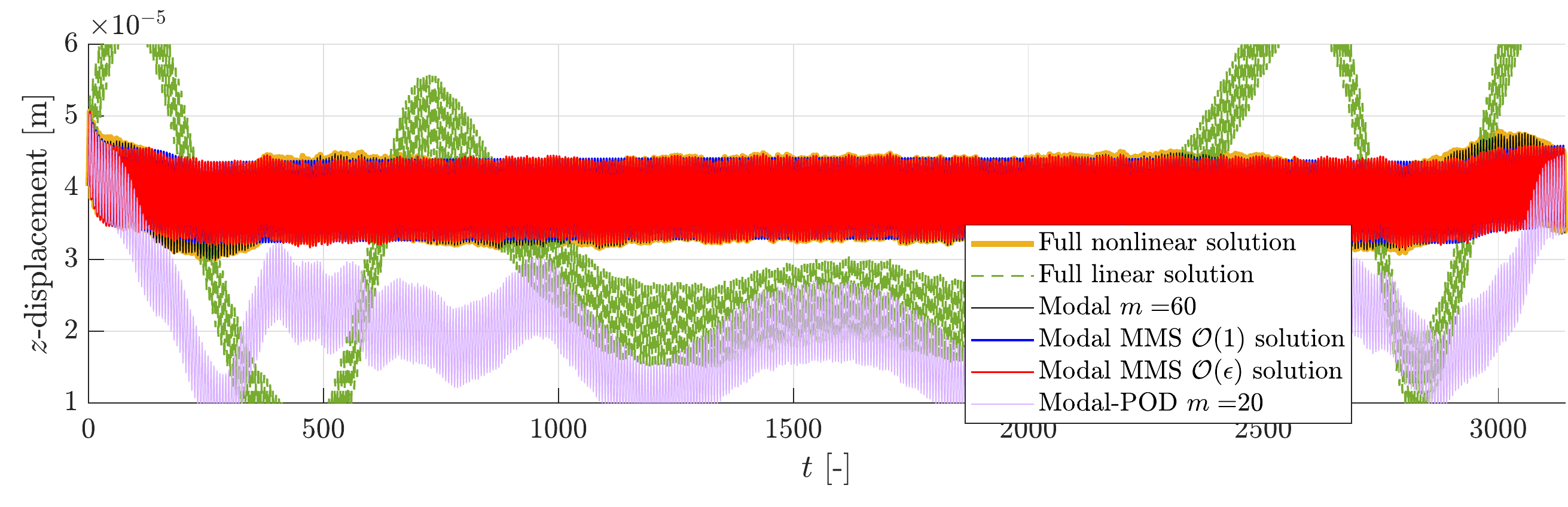}
		\caption{}
	\end{subfigure}
	\begin{subfigure}{\textwidth}
		\centering\includegraphics[width=\linewidth]{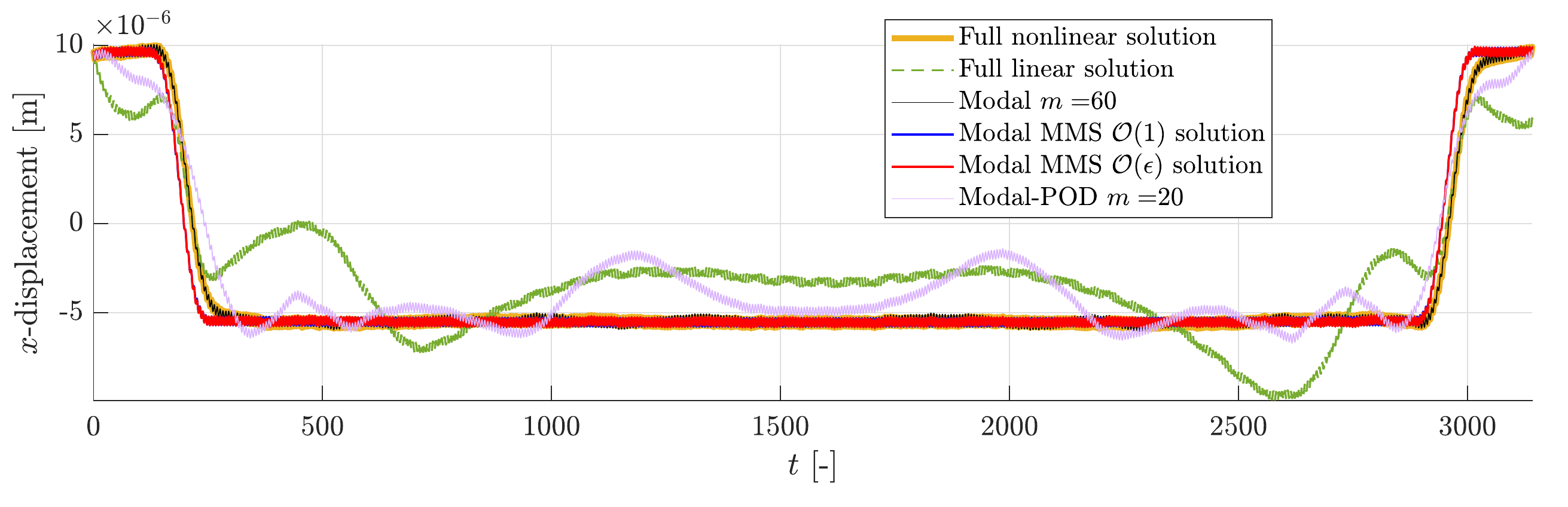}
		\caption{}
	\end{subfigure}
	\begin{subfigure}{\textwidth}
		\centering\includegraphics[width=\linewidth]{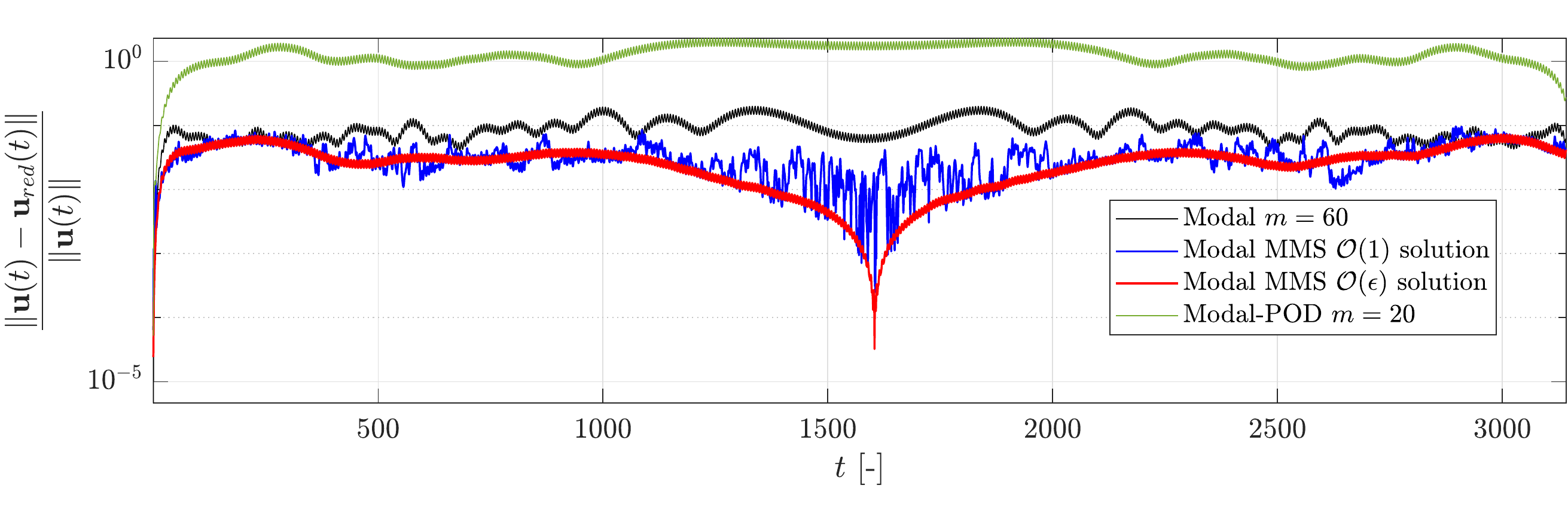}
		\caption{}
	\end{subfigure}
	
	\caption{ \label{fig:curved_nlin} \small The (a) $ z$-direction and (b) $ x$-direction displacements for the node situated at $ x=L/4 $ in response to combined thermo-mechanical loading \eqref{eq:mech_load} (cf.~Figure~\ref{fig:randper}), \eqref{eq:thermal_dynamics} with $ x_0 = 0.1L, A=0.8L, \epsilon=10^{-3} $. The Full system solution is depicted in the solid grey line; the broken green line shows the linearized response that is desirably different from the nonlinear one; the solid black line (Modal $ m=60 $) shows the reduced solution obtained from Galerkin projection onto a constant basis of rank 60; the solid blue (Modal MMS $ \mathcal{O}(1) $) and red (Modal MMS $ \mathcal{O}(\epsilon) $) lines show the reduced solution using the MMS-based reduction upto $ \mathcal{O}(1)  $ and $ \mathcal{O}(\epsilon) $ accuracy respectively, using a basis of size $ M=20 $  adapting to the instantaneous temperature configuration. The instantaneous error~\eqref{eq:error_t} of the different reduced solutions relative to the full nonlinear solution is shown in (c).  }
\end{figure}

To compare the results with another reduction approach, we again use the stacking-based approach in the following two ways: 

\begin{enumerate}
\item Modal ($ m=60 $): Similar to the linear case, we again use a subset of randomly selected temperature configuration ($ x_c^{(j)}, j=4,7,13 $) from the database. In contrast to the linear case, here we obtain a reduction basis of size 60 (since each of the three temperature configurations correspond to a basis of size 20).
\item Modal-POD ($ m=20 $) approach, where we use the $ m=20 $ modes associated to the highest singular values of the matrix $ \boldsymbol{\mathcal{V}} $ which contains the first five VMs and the corresponding 15 modal derivatives at $ n_d = 19 $ temperature configurations (cf.~eq.~\eqref{eq:temp_configs},\eqref{eq:Vdef}).
\end{enumerate}
 Figures~\ref{fig:curved_nlin}(a) and (b) show the $ z $ and $ x $-direction displacements at the quarter span of the beam for $ \epsilon=0.001 $. To provide a more global picture, we compare the error in the vector of generalized displacements relative to the full nonlinear solution $ \mathbf{u}(t) $ at each time instant (cf. Figure~\ref{fig:curved_nlin}c). This error is defined as\begin{equation}
\label{eq:error_t}
\mathbf{e}_{red}(t):= \frac{\|\mathbf{u}(t)-\mathbf{u}_{red}(t)\|_2}{\|\mathbf{u}(t)\|_2}\,
\end{equation}
where $ \mathbf{u}_{red}(t) $ denotes the reduced solution obtain from the different approaches. It is interesting that despite have a basis size of 60, a constant basis reduction provides a worse accuracy than the MMS-based reduction with a basis of size 20 that adapts to the instantaneous temperature configuration.


The improvement in accuracy between the $ \mathcal{O}(1) $ and the $ \mathcal{O}(\epsilon) $ reduced solutions from the MMS  is not very apparent from Figure~\ref{fig:curved_nlin}c. For this reason, we use the following global error estimate which is uniform over time interval $ [0, I] $, where $ I $ is the total time over which the simulation is performed
\begin{equation}
\label{eq:E}
\mathbf{E}_{red} := \frac{\int_0^I \|\mathbf{u}(t)-\mathbf{u}_{red}(t)\|_2~ \mathrm{d}t}{\int_0^I \|\mathbf{u}(t)\|_2~\mathrm{d}t} \approx \frac{\sum_{t\in\mathcal{I}} \|\mathbf{u}(t)-\mathbf{u}_{red}(t)\|_2}{\sum_{t\in\mathcal{I}} \|\mathbf{u}(t)\|_2},
\end{equation}
where the approximation is perform using a one-point numerical integration over the uniformly sampled (over $ [0,I] $) time instants in the set $ \mathcal{I} $. According to Table~\ref{tab:error}, this estimate of error shows a marginal improvement in accuracy using the $ \mathcal{O}(\epsilon) $ solution over the $ \mathcal{O}(1) $ solution, as expected.
\begin{table}[H]
	\begin{centering}
		\begin{tabular}{l c r}
			\hline 
			Reduction method  & Basis size ($ m $) & Error \tabularnewline
			\hline 
			\hline 
			Modal  & 60 &9.47\% \tabularnewline
			Modal-POD & 20 & 136.75\%\tabularnewline
			Modal MMS $ \mathcal{O}(1) $ & 20  & 3.69\% \tabularnewline
			Modal MMS $ \mathcal{O}(\epsilon) $ & 20  & 3.19\% \tabularnewline
			\hline 
		\end{tabular}
		\par\end{centering}
	\caption{\label{tab:error} \small The uniform-in-time error estimate~\eqref{eq:E} for the different reduction approaches considered in the curved beam with geometric nonlinearities (cf.~Figure~\ref{fig:curved_nlin}).}
\end{table}
\section{Discussion}
\label{sec:disc}
The numerical examples above serve as proof of concept for effective reduction using the MMS-based approach, proposed in this work. Although these examples are considerably simpler than the targeted applications of model reduction techniques, we do not envision issues with the scalability of this systematic approach. This is because the adaptive interpolation ensures that the number of reduced variables remains uniform as the temperature is dynamically changing during a simulation. Nonetheless, we discuss some general sources of computational bottleneck and possible ways to address them in the following.

\paragraph{Arbitrarily varying temperature fields.} For simplicity, we have only considered externally-imposed temperature variation on the beam in this work. However, this variation in structural systems is usually obtained by solving the heat equation~\eqref{eq:heat}, which is given in its linearized form by the system
\begin{equation}
\label{eq:lin_heat}
\mathbf{M}_T\mathbf{T}^{\prime} + \mathbf{K}_T\mathbf{T} = \mathbf{h}(\tau),
\end{equation}
where the heat source $ \mathbf{h} $ is externally applied, and the temperature $ \mathbf{T} $ is indeed obtained as an output. 
For the numerical examples treated in this work, a single parameter ($ x_c $) was sufficient to model the variation in temperature distribution over the beam. Thus, the database of reduction bases (used in adaptive basis selection) was obtained for a range of single-parameter values~\eqref{eq:temp_configs}. General temperatures fields, however, require $ n_T $ (number of DOFs in eq.~\eqref{eq:heat}) parameters for description, which could be potentially large and would require a huge database of reduction bases to be computed. Thus, the memory requirements for the adaptive basis interpolation could be a potential bottleneck for adaptive basis selection in case of multi-dimensional parameters. One simple approach to reduce the number of temperature field parameters is as follows.

The temperature distribution of the structure can be spectrally
decomposed into a basis of thermal modes obtained from the eigenvalue analysis of the linear operators in $\eqref{eq:lin_heat}$ as 
\begin{equation}
\left(\mathbf{K}_T+\lambda_{j}\mathbf{M}_{T}\right)\boldsymbol{\psi}_{j}=\mathbf{0},\quad j\in\{1,2,\dots,n_{T}\}
\end{equation}
where  $\mathbf{\boldsymbol{\psi}}_{j}$
is the $ j ^{\mathrm{th}}$ eigenvector related to the thermal problem ($\ref{eq:heat}$)
and $\lambda_{j}$ is the corresponding eigenvalue. The temperature configuration can then be suitably approximated using a few thermal modes as
\begin{equation}
\mathbf{T}(\tau)=\sum_{j=1}^{n_{T}}\boldsymbol{\psi}_{j}q_{T,j}(\tau)\approx\boldsymbol{\Psi}\,\mathbf{q}_{T}(\tau),
\end{equation}
where $\boldsymbol{\Psi}\in\mathbb{R}^{n_{T}\times m_{T}}$ is a matrix
containing the thermal modes of the structure and the slowly varying
amplitudes $\mathbf{q}_{T}(\tau)\in\mathbb{R}^{m_{T}}$ of the thermal
modes determine the temperature distribution across the structure
at a given time. It is then easy to see that the reduction basis for structural dynamics can be parameterized by the thermal mode amplitudes $\mathbf{q}_{T}$, thereby reducing the number of required parameters to $ m_T\ll n_T $. 

\paragraph{Computational time and hyper-reduction.}
In the case of linear systems such as eq.~\eqref{eq:lin} with time-invariant coefficient matrices, the reduced operators can be precomputed at once when a constant reduction basis is used. However, with time-dependence in the coefficient matrices such as in eq.~\eqref{eq:lin_temp}), the reduced operators need to be updated online at each time step during time integration. This operation poses computational bottlenecks in reduction of such systems, and a reduction in the number of unknowns does not lead to a tantamount reduction in computational time. Note that this issue affects reduction techniques using a constant basis, as well as the MMS-based reduction technique proposed here. A possible solution to this issue is to precompute the reduced operators $\mathbf{M}_{\boldsymbol{\Phi}}$, $\mathbf{C}_{\boldsymbol{\Phi}}$ and $\mathbf{K}_{\boldsymbol{\Phi}}$ in eq.~\eqref{eq:linO1} along with the reduction bases at each temperature configuration in the database, during the offline stage. Thereafter, this reduced information can be directly interpolated online (cf.~ref.~\cite{Amsallem}) during time integration. This completely avoids the interpolation of reduction bases and projection at each time step at the cost of precomputation of reduced information.

For nonlinear reduced-order models, the evaluation of reduced nonlinear operators by projection onto the reduction basis can also be a computational bottleneck~\cite{DEIM} during online time integration. Indeed, due to these extra operations, the computation of a reduced solution becomes possibly more expensive than that for a full system solution~\cite{MasterThesis}. Hyper-reduction techniques provide reprieve in such cases by a fast approximation of the reduced nonlinear operators using training and selection procedures. In the finite-element context, the energy conserving sampling and weighing method~(ECSW)~\cite{ECSW1} is found to be effective. Based on offline training, this method provides a single reduced finite-element mesh for a given set of reduction bases and parameter values, as also discussed in ref.~\cite{Chapmann}. Furthermore, the hyper-reduction of geometrically nonlinear structural systems using ECSW with VMs and modal derivatives basis has already been affirmed in previous works~\cite{Jain_2018,Jain_2019}. These features make it possible to train a ROM across a range of temperature configurations and obtain a reduced mesh so as to accelerate computation for adaptive basis reduction using the MMS-based approach proposed in this work.

\section{Conclusion}
\label{sec:conc}
In this work, we focused on model reduction for temperature-dependent structural dynamics equations with time varying temperature fields. We discussed that a slow variation in the structural temperature is not only physically relevant but also essential to justify model reduction using an adaptive reduction basis (cf. Section~\ref{sec:introduction}). Subsequently, we proposed the systematic use of the method of multiple scales to exploit this slow temperature dependence in the structural dynamics equations. We consistently reduced the equations of motion using a temperature-dependent basis that slowly adapts to the instantaneous temperature configuration of the structure (cf.~Section~\ref{sec:basis}). 

We treated numerical examples in the linear case using an adaptive basis containing vibration modes and in the geometrically nonlinear setting, the basis was enriched using the corresponding modal derivatives. In the process, we also concluded (cf.~Section~\ref{sec:results}) that the use constant reduction basis obtained by systematically combining reduction bases obtained from different temperature configurations is not a scalable strategy, as this results in a potentially large number of reduced variables. On the other hand, the MMS-based reduction approach turned out to be very efficient in terms of number of variables and showed consistently better accuracy. 

While we addressed general computational bottlenecks in the reduction procedure and discussed possible strategies to mitigate them (cf.~Section~\ref{sec:disc} ), the main focus of this work was efficient dimensionality reduction of the governing equations. Implementation of these strategies in line with the discussion in Section~\ref{sec:disc} is essential to obtain the necessary and sufficient computational speed. This forms part of our future efforts. In combination with these measures, we expect the MMS-based adaptive reduction would enable us to reduce realistic structures under the influence of thermal environments in a robust and efficient manner.

\paragraph{Acknowledgments.}
The authors would like to thank the anonymous reviewers of this work for their very instrumental feedback. The support of the Air Force Office of Scientific Research, Air Force
Material Command, USAF under Award No.FA9550-16-1-0096 is acknowledged.


\begin{thebibliography}{10}
\bibitem{Rixen97}G{\'e}radin, M., Rixen, D. Mechanical Vibrations:
Theory and Application to Structural Dynamics, 2nd Edition, \textit{Wiley}
(1997). ISBN: 0-471-97524-9


\bibitem{IC} Idelsohn, S. R., Cardona, A. A reduction method for
nonlinear structural dynamic analysis, \emph{Comput Methods Appl Mech
Eng} (1985) \textbf{49}(3): 253-279. DOI: \href{https://doi.org/10.1016/0045-7825(85)90125-2}{10.1016/0045-7825(85)90125-2}


\bibitem{POD1}Kosambi, D. Statistics in function space,\emph{ \href{http://repository.ias.ac.in/99240/}{J. Indian Math. Soc.}}
(1943) \textbf{7}, 76\textendash 78. 

\bibitem{POD2}Amabili, M., Sarkar, A., Paidoussis, M.P. Reduced-order
models for nonlinear vibrations of cylindrical shells via the proper
orthogonal decomposition method. \emph{J. Fluids Struct. }(2003) \textbf{18}(2):227-250.
DOI: \href{https://doi.org/10.1016/j.jfluidstructs.2003.06.002}{10.1016/j.jfluidstructs.2003.06.002} 

\bibitem{Hollkamp_ICE}Hollkamp, J.J. and Gordon, R.W., Reduced-order
models for nonlinear response prediction: Implicit condensation and
expansion, \emph{J. Sound Vib. }(2008) \textbf{318}: 1139-1153. DOI:
\href{https://doi.org/10.1016/j.jsv.2008.04.035}{10.1016/j.jsv.2008.04.035}

\bibitem{mignolet}Mignolet, M.P., Przekop, A., Rizzi, S.A., Spottswood,
S.M. A review of indirect/non-intrusive reduced order modeling of
nonlinear geometric structures. \textit{J Sound Vib} (2013) 332(10):
2437-2460. DOI: \href{https://doi.org/10.1016/j.jsv.2012.10.017}{10.1016/j.jsv.2012.10.017}


\bibitem{qm}Jain, S., Tiso, P., Rixen, D.J., Rutzmoser, J.B. A Quadratic
Manifold for Model Order Reduction of Nonlinear Structural Dynamics,
\textit{Comput. Struct.} (2017), 188: 80-94. DOI: \href{https://doi.org/10.1016/j.compstruc.2017.04.005}{10.1016/j.compstruc.2017.04.005}


\bibitem{Przekop_2007}Przekop, A. and Rizzi, S.A., Dynamic Snap-Through
of Thin-Walled Structures by a Reduced-Order Method, \emph{AIAA Journal
}(2007) \textbf{45}(10): 2510-2519. DOI: \href{https://doi.org/10.2514/1.26351}{10.2514/1.26351}

\bibitem{Spottswood_2010}Spottswood, S.M., Hollkamp, J.J. and Eason,
T.G., Reduced-Order Models for a Shallow Curved Beam Under Combined
Loading, \emph{AIAA Journal} (2010) \textbf{48}(1): 47-55. DOI: \href{https://doi.org/10.2514/1.38707}{10.2514/1.38707}

\bibitem{Amsallem}Amsallem, D. and Farhat, C. An online method for
interpolating linear parametric reduced-order models, \emph{SIAM J.
Sci. Comput.}, (2011) \textbf{33}(5): 2169-2198. DOI: \href{https://doi.org/10.1137/100813051}{10.1137/100813051} 

\bibitem{Amsallem_2009}Amsallem, D., Cortial, J., Carlberg, K. and
Farhat, C. A method for interpolating on manifolds structural dynamics
reduced-order models, \emph{Int. J. Numer. Meth. Engng} 2009; \textbf{80}:1241\textendash 1258.
DOI: \href{https://doi.org/10.1002/nme.2681}{10.1002/nme.2681}

\bibitem{Lieu_2004} Lieu, T. and Lesoinne, M. Parameter Adaptation of Reduced Order Models for Three-Dimensional Flutter Analysis, \emph{42nd AIAA Aerospace Sciences Meeting and Exhibit}, Reno, Nevada, 5-8 Jan,  2004, \textbf{80}:1241\textendash 1258.
DOI: \href{https://doi.org/10.2514/6.2004-888}{10.2514/6.2004-888}


\bibitem{Falk_2011} Falkiewicz, N. J., Cesnik, C. E. S., Crowell,
A. R., McNamara, J. J. Reduced-Order Aerothermoelastic Framework for
Hypersonic Vehicle Control Simulation, \emph{AIAA Journal} (2011)\textbf{
	49}(8): 1625-1646. DOI: \href{https://doi.org/10.2514/1.J050802}{10.2514/1.J050802} 

\bibitem{Matney_2014}Matney, A., Reduced Order Model-Based Prediction
of the Nonlinear Geometric Response of a Panel Under Thermal, Aerodynamic,
and Acoustic Loads, \emph{\href{http://hdl.handle.net/2286/R.A.143484}{PhD Thesis}},
Arizona State University (2014). 

\bibitem{Perez_2011}Perez, R., Wang, X.Q. and Mignolet, M.P., Nonlinear Reduced-Order Models for Thermoelastodynamic
Response of Isotropic and Functionally
Graded Panels, \emph{AIAA Journal} (2011) \textbf{49}(3): 630-641. DOI: \href{https://doi.org/10.2514/1.J050684}{10.2514/1.J050684}


\bibitem{Tamarozzi_2014}Tamarozzi, T., Heirman, G.H.K., Desmet, W.,
An on-line time dependent parametric model order reduction scheme
with focus on dynamic stress recovery, \emph{Comput Methods Appl Mech
	Eng} (2014) \textbf{268}: 336-358. DOI: \href{https://doi.org/10.1016/j.cma.2013.09.021}{10.1016/j.cma.2013.09.021}

\bibitem{Kevorkian}Kevorkian, J. and Cole, J. D. Multiple Scale and
Singular Perturbation Methods \emph{Springer}, New York(1996), Vol.
114, ed: JE. Marsden L. Sirovich F. John. ISBN: 978-1-4612-8452-9. DOI:\href{https://doi.org/10.1007/978-1-4612-3968-0}{10.1007/978-1-4612-3968-0}













\bibitem{IC2} Idelsohn, S. R., Cardona, A. A load-dependent basis for reduced nonlinear structural dynamics, \emph{Comput Struct} (1985) \textbf{20}(1-3): 203-210. DOI: \href{https://doi.org/10.1016/0045-7949(85)90069-0}{10.1016/0045-7949(85)90069-0}


\bibitem{qm2}Rutzmoser, J.B., Rixen, D.J., Tiso, P., Jain, S.  Generalization of quadratic manifolds for reduced order modeling of nonlinear structural dynamics, \textit{Comput. Struct.} (2017), \textbf{192}: 196-209. DOI: \href{https://doi.org/10.1016/j.compstruc.2017.06.003}{10.1016/j.compstruc.2017.06.003}

%

\bibitem{WWS}Weeger, O., Wever, U., Simeon, B.
On the use of modal derivatives for nonlinear model order reduction, \emph{Int. J. Numer. Meth. Engng} (2016) \textbf{108}: 1579--1602. DOI: \href{http://dx.doi.org/10.1002/nme.5267}{10.1002/nme.5267} 









%


%
%
%
%
%
%
%

\bibitem{DEIM}Chaturantabut, S., Sorensen, D. C. Nonlinear model
reduction via discrete empirical interpolation, \textit{SIAM Journal
	on Scientic Computing} (2010) \textbf{32}(5): 2737-2764. DOI: \href{https://doi.org/10.1137/090766498}{10.1137/090766498}

\bibitem{MasterThesis}Jain, S. Model order reduction for nonlinear
structural dynamics, Delft University of Technology (2015) \emph{Master's
	Thesis. }TU Delft repository UUID: \href{http://resolver.tudelft.nl/uuid:cb1d7058-2cfa-439a-bb2f-22a6b0e5bb2a}{cb1d7058-2cfa-439a-bb2f-22a6b0e5bb2a}

\bibitem{ECSW1}Farhat, C., Avery, P., Chapman, T., Cortial, J. Dimensional
reduction of nonlinear finite element dynamic models with finite rotations
and energy-based mesh sampling and weighting for computational efficiency,
\textit{Int. J. Numer. Meth. Engng} (2014) \textbf{98}(9): 625--662. DOI: \href{https://doi.org/10.1002/nme.4668}{10.1002/nme.4668}

\bibitem{Chapmann} Chapman, T., Avery, P., Collins, P. and Farhat, C. Accelerated mesh sampling for the hyper reduction of nonlinear computational models,
\textit{Int. J. Numer. Meth. Engng} (2017) \textbf{109}:1623--1654. DOI: \href{https://doi.org/10.1002/nme.5332}{10.1002/nme.5332}

\bibitem{Jain_2018} Jain, S., Tiso, P. Simulation-free hyper-reduction for geometrically nonlinear structural dynamics: A quadratic manifold lifting approach. \emph{J. Comput. Nonlinear Dynam.} (2018) 13(7), 071003. DOI: \href{https://doi.org/10.1115/1.4040021}{10.1115/1.4040021}

\bibitem{Jain_2019} Jain, S., Tiso, P. Hyper-Reduction Over Nonlinear Manifolds for Large Nonlinear Mechanical Systems. \emph{J. Comput. Nonlinear Dynam.} (2019) 14(8): 081008. DOI: \href{https://doi.org/10.1115/1.4043450}{10.1115/1.4043450}


\bibitem{Golub2013} Golub, G. H. \& van Loan, C. F. Matrix Computations, The Johns Hopkins University Press (2013), Baltimore, ISBN:9781421407944.

\bibitem{Tiso} Tiso, P. Optimal second order reduction basis selection for nonlinear transient analysis. In T. Proulx (Ed.), Modal Analysis Topics, Volume 3. Conference Proceedings of the Society for Experimental Mechanics Series (pp. 27?39). Springer, New York, NY  (2011). DOI: \href{https://doi.org/10.1007/978-1-4419-9299-4_3}{10.1007/978-1-4419-9299-4\_3}
\end{thebibliography}
\end{document}